# Verifying Solutions to Semantics-Guided Synthesis Problems

CHARLIE MURPHY, University of Wisconsin—Madison, USA*
KEITH JOHNSON, University of Wisconsin—Madison, USA
THOMAS REPS, University of Wisconsin—Madison, USA
LORIS D'ANTONI, University of California—San Diego, USA

Semantics-Guided Synthesis (SemGuS) provides a framework to specify synthesis problems in a solver-agnostic and domain-agnostic way, by allowing a user to provide both the syntax and semantics of the language in which the desired program should be synthesized. Because synthesis and verification are closely intertwined, the SemGuS framework raises the following question: *how does one verify that a user-given program satisfies a given specification when interpreted according to a user-given semantics?*

In this paper, we prove that this form of *language-agnostic* verification (specifically that verifying whether a program is a valid solution to a SemGuS problem) can be reduced to proving validity of a query in the $\mu$CLP calculus, a fixed-point logic that is capable of expressing alternating least and greatest fixed-points. Our encoding into $\mu$CLP allows us to further classify the SemGuS verification problems into ones that are reducible to satisfiability of (*i*) first-order-logic formulas, (*ii*) Constrained Horn Clauses, and (*iii*) $\mu$CLP queries. Furthermore, our encoding shines light on some limitations of the SemGuS framework, such as its inability to model nondeterminism and reactive synthesis. We thus propose a modification to SemGuS that makes it more expressive, and for which verifying solutions is exactly equivalent to proving validity of a query in the $\mu$CLP calculus. Our implementation of SemGuS verifiers based on the above encoding can verify instances that were not even encodable in previous work. Furthermore, we use our SemGuS verifiers within an enumeration-based SemGuS solver to correctly synthesize solutions to SemGuS problems that no previous SemGuS synthesizer could solve.

## 1 INTRODUCTION

In program synthesis, the goal is to find a program in a given search space that meets a given specification. Synthesis has found great successes in specific domains, e.g., spreadsheet transformations [36] and bit-vector manipulations [20], where the search space is fixed and its properties can be exploited to design powerful domain-specific synthesis solvers. However, for synthesis to become a general-purpose technology that can help users with a variety of tasks, one should be able to customize the search space and specifications of a synthesis problem in a programmable way that is agnostic of a specific domain or synthesis solver.

To address the problem of making synthesis "programmable", Kim et al. [28] proposed the SemGuS framework, which enables one to specify synthesis problems in a solver-agnostic and domain-agnostic way. The key differentiating aspect of the SemGuS framework is that a user provides a set of Constrained Horn Clauses (CHCs) whose *least solution* defines the semantics of the programming language over which one is interested in performing synthesis. (A detailed example of a SemGuS problem is given in Figure 1 and discussed in §2.1.) While this formalism enables a great deal of flexibility when describing a synthesis problem—e.g., one can naturally define the operational semantics of an imperative programming language—this generality comes at a cost: building solvers for general SemGuS problems can be difficult [15].

---

*This work submitted prior to joining Amazon Web Services.

Due to this complexity, most of the work on SemGuS has limited its focus to problems whose specifications are given as a finite set of examples [28]. To go beyond specifications with examples and consider specifications with quantified variables, we require a technique capable of verifying whether a candidate solution is correct. Verification is not only needed to check that the final solution meets the specification, it is also often used to implement synthesis algorithms that use enumeration and constraint-solving to efficiently explore the search space of possible programs—i.e., *verifying whether a candidate solution is correct is a crucial missing component that is needed for solving SemGuS problems involving complex specifications, and for building effective SemGuS solvers.*

One of the key challenges raised by SemGuS (and the focus of this paper) is *how does one build a language-agnostic verifier that can check if a program matches a specification?*—i.e., when the verification user provides the program, specification, *and* the semantics of the language in which the program should be interpreted. Specifically, in SemGuS the user provides the semantics of the programming language as a set of CHCs [8]—i.e., a set of Horn clauses augmented with first-order theories that give meaning to a set of relations defining the semantics of programs.

From the standpoint of the SemGuS framework, the language for which we are performing verification is *not* fixed—and can in fact be arbitrary—and the verification technique needs to be able to reason about the CHCs that define a language's semantics. In particular, because the semantics of the input language is provided logically, there is no easy way to relate it to known verification approaches that are tailored to specific programming constructs. This last aspect makes existing verification approaches that are tied to specific programming languages [21, 24, 31, 35, 43] not suitable for verifying solutions to SemGuS problems. In particular, these verification approaches take advantage of a fixed programming language and its fixed semantics to use specialized techniques such as loop-invariants and Hoare-style reasoning for imperative programs [31] and type-based reasoning for functional programs [35]. The recently released SemGuS Toolkit [26] contains a baseline verifier as part of their KS2 SemGuS synthesizer. While the baseline verifier is in theory capable of handling specifications beyond examples, it does so via a naive encoding of the specification and CHC semantic relations as an SMT formula with recursively defined functions, a theory that SMT solvers can rarely reason about successfully.

In this paper, we present a comprehensive study of the problem of verifying solutions to SemGuS problems. The first contribution of this paper is the following: given a program $p$, a semantics defined using Constrained Horn Clauses *Sem*, and a specification $\varphi$ (which is allowed to mention the semantic relations defined by *Sem*), we show that the problem of verifying whether $p$—when evaluated according to *Sem*—satisfies $\varphi$ can be expressed as a validity check in the $\mu$CLP calculus [40]. $\mu$CLP is a fixed-point logic that generalizes CHCs and co-CHCs by combining both least and greatest fixed-points with interpreted first-order theories. While SemGuS uses only least fixed-points to define the semantics of programs (i.e., as the least solution to a set of CHCs), the fact that the semantic relations can appear in both positive and negative positions in a user-supplied specification means that reasoning needs to be carried out in a proper fixed-point logic, such as $\mu$CLP. Because of the complexity of building solvers for checking validity of $\mu$CLP queries, the second contribution of the paper is to identify fragments of SemGuS verification problems that can be reduced to checking satisfiability of first-order-logic formulas (Satisfiability Modulo

Theories) and CHCs, for which more scalable solvers exist [10, 22]. Furthermore, because we show an equivalence between SemGuS verification and $\mu$CLP satisfiability (Theorem 4.8), these reductions are directly applicable to checking satisfiability of $\mu$CLP formulas themselves.

Our study highlights a strong connection between SemGuS and $\mu$CLP, and raises the question of whether there exist synthesis problems for which verification is expressible using $\mu$CLP, but cannot be expressed in SemGuS. We answer this question affirmatively by showing that SemGuS cannot reason about programs involving nondeterminism and games, both of which are commonly found in reactive-synthesis problems [3]. To close the loop between $\mu$CLP and SemGuS, we define a minimal extension of SemGuS—i.e., we allow relations to appear in a negated form in the semantic definitions and require an ordering over semantic relations—which results in a new framework that aligns exactly with what is verifiable using $\mu$CLP. Finally, we incorporate our verification technique into a synthesizer for SemGuS problems that is capable of solving SemGuS problems with complex logical specifications.

*Contributions.* Our work makes the following contributions.

- We identify how the problem of verifying programs in SemGuS is tightly related to proving validity in fixed-point logics (§2).
- We propose an extension of the SemGuS framework that can capture, e.g., reactive synthesis problems (§3).
- We analyze when solutions to SemGuS problems can be verified using various logical fragments (SMT, CHC, and $\mu$CLP) and show that our extension of SemGuS aligns exactly with what is verifiable using $\mu$CLP (§4) .
- We implemented our approach in a tool, Muse, together with several optimizations (§5), and used Muse to verify (or disprove correctness for) solutions to SemGuS problems (§6). The problems comprise a variety of functional and reactive-synthesis problems, encoding such domains as imperative programs, regular expressions, Büchi games, and robot path planning.
- We incorporated Muse within the SemGuS synthesizer Ks2 to enable Ks2 to solve SemGuS problems that could not be solved by any previous approach (§6). In particular, by incorporating Muse into Ks2, we enabled Ks2 to solve 135 SemGuS problems involving specifications with quantified variables (74 of which could not be solved by Ks2 using the baseline verifier).

§7 discusses related work. §8 concludes.

## 2 OVERVIEW

This section illustrates our approaches for verifying a solution to a SemGuS problem using three problems of increasing complexity. Our technique reduces the verification task to checking validity of queries in various logical fragments that can be dispatched to existing solvers. The examples should provide enough details to understand the SemGuS framework.

### 2.1 Max2: Quantified SMT

Consider the problem of synthesizing a loop-free imperative program with two variables x and y that computes the maximum of two values. Figure 1 gives all the components necessary to define this synthesis problem in the SemGuS framework:

- A grammar $G_{max2}$ defining the syntax of the language under consideration (Figure 1a).
- A set of constrained Horn clauses $Sem_{max2}$ that inductively define (as the least solution of $Sem_{max2}$) the semantics of all programs in the language (Figure 1d).
- A specification $\varphi_{max2}$ that describes how the synthesized program should behave when evaluated according to $Sem_{max2}$ (Figure 1b).

$S ::= \texttt{x = } E \mid \texttt{y = } E$
$\mid S;S \mid \texttt{Ite } B\ S\ S$

$E ::= \texttt{0} \mid \texttt{1} \mid \texttt{x} \mid \texttt{y} \mid E\texttt{+}E$

$B ::= E \texttt{ < } E$

(a) Grammar $G_{max2}$

$$\forall x, y, x'. \begin{pmatrix} \exists y'. Sem_S(max2, x, y, x', y') \\ \Updownarrow \\ (x' = x \lor x' = y) \land x \leq x' \land y \leq x' \end{pmatrix}$$

(b) Specification $\varphi_{max2}$

$\texttt{Ite}$
$\quad(\texttt{x < y})$
$\quad(\texttt{x = y})$
$\quad(\texttt{x = x})$

(c) Solution $s_{max2}$

$$\frac{Sem_E(e, x, y, x') \land y = y'}{Sem_S(\texttt{x = e}, x, y, x', y')} \qquad \frac{Sem_E(e, x, y, y') \land x = x'}{Sem_S(\texttt{y = e}, x, y, x', y')} \qquad \frac{Sem_B(b, x, y, \top) \quad Sem_S(\texttt{t}, x, y, x', y')}{Sem_S(\texttt{Ite b t e}, x, y, x', y')}$$

$$\frac{Sem_S(\texttt{s}, x, y, x'', y'') \quad Sem_S(\texttt{t}, x'', y'', x', y')}{Sem_S(\texttt{s; t}, x, y, x', y')} \qquad \frac{Sem_B(b, x, y, \bot) \quad Sem_S(\texttt{e}, x, y, x', y')}{Sem_S(\texttt{Ite b t e}, x, y, x', y')}$$

$$\frac{}{Sem_E(\texttt{0}, x, y, 0)} \qquad \frac{}{Sem_E(\texttt{1}, x, y, 1)} \qquad \frac{Sem_E(\texttt{s}, x, y, r_s) \quad Sem_E(\texttt{t}, x, y, r_t) \quad b \Leftrightarrow r_s < r_t}{Sem_B(\texttt{s < t}, x, y, b)}$$

$$\frac{Sem_E(\texttt{s}, x, y, r_s) \quad Sem_E(\texttt{t}, x, y, r_t) \quad r = r_s + r_t}{Sem_B(\texttt{s + t}, x, y, r)} \qquad \frac{}{Sem_E(\texttt{x}, x, y, x)} \qquad \frac{}{Sem_E(\texttt{y}, x, y, y)}$$

(d) Semantics $Sem_{max2}$

Fig. 1. SemGuS definition for the problem of synthesizing an *imperative program max2* that computes the maximum of two input values $x$ and $y$. Figure 1a contains a regular tree grammar defining the syntax of the language we can use to build programs (i.e., imperative programs with if-then-else, comparisons, and linear assignments). The semantics of the language is inductively defined using constrained Horn clauses (Figure 1d)—e.g., the semantics of programs derivable from nonterminal $S$ is given via the inductively defined relation $Sem_S(s, x, y, x', y')$ where, for example, $Sem_S(\texttt{x=1}, 3, 3, 1, 3)$ denotes that running the program x=1 with initial values of 3 for both $x$ and $y$ results in a state where $x$ is 1 and $y$ is 3. Figure 1b specifies when the solution is correct: on an input state $x, y$, the program $max2$ should output a state $x', y'$ such that $x'$ is greater or equal than the values of $x$ and $y$ and is equal to one of them. The program in Figure 1c (parenthesis are added for readability) is a possible solution to this SemGuS problem—this program is in the grammar $G_{max2}$ and when evaluated on any possible input state according to the semantics $Sem_{max2}$, it satisfies the specification $\varphi_{max2}$.

The specification $\varphi_{max2}$ states that the synthesized program (represented symbolically by the variable $max2$) must terminate in a state in which $x'$ (i.e., the final value of the variable x) is the maximum of the initial-state values assigned to variables x and y—i.e., $x$ and $y$. Solving this SemGuS problem means providing a program in the grammar that satisfies this specification when evaluated according to the semantics.

Figure 1c presents a *candidate* solution $s_{max2}$ to this SemGuS problem. Rather than determining how to synthesize $s_{max2}$, this paper tackles the following question: how do we show that when the program $s_{max2}$ is "evaluated" according to the semantics $Sem_{max2}$, it satisfies the specification $\varphi_{max2}$.

*Beyond CHCs.* Because the semantics is already defined as the least solution to a set of CHCs, it is natural to expect to be able to solve the problem in terms of CHC satisfiability. That is, at first blush, it seems plausible to check the query in Equation (1), which states that $\varphi_{max2}$ is valid when interpreted using the least solution of the semantic relations ($Sem_{max2}^{LFP}$).

$$Sem_{max2}^{LFP} \models \forall x, y, x'.(\exists y'.Sem_S(max2, x, y, x', y') \Leftrightarrow (x' = x \lor x' = y) \land x \leq x' \land y \leq x') \quad (1)$$

While the semantics is defined as the least solution to a set of constrained Horn clauses $Sem_{max2}$, the positive occurrence of $Sem_S$ within $\varphi_{max2}$ results in a query that cannot be reasoned about using the typical decision procedure for CHCs (i.e., CHC satisfiability). Given a set of CHCs $C$ and a query $\psi$, the typical approach to prove $\psi$ is valid under the least solution of $C$ is to instead check if $C \land \neg \psi$ is unsatisfiable. This translation is sound provided that $\psi$ contains only negative occurrences of the uninterpreted relations defined by $C$. Because the specification $\varphi_{max2}$ contains a positive occurrence of $Sem_S$, we must turn to a different approach to solving the validity query.

*Finite derivation trees can be desugared.* Our first insight is that for problems like $max2$, where the semantic definitions are recursively defined with respect to the term's proper subterms, one can always build a finite derivation tree that describes the semantics of a given program. For example, the derivation tree for $s_{max2}$ is as follows:

$$\frac{\dfrac{r_s = x}{Sem_E(x,x,y,r_s)} \quad \dfrac{r_t = x}{Sem_E(y,x,y,r_t)} \quad r_s < r_t}{Sem_B(x < y, x, y, b)} \quad b \wedge \dfrac{\dfrac{y = x'}{Sem_E(y,x,y,x')} \quad y = y'}{Sem_S(x = y, x, y, x', y')} \vee \neg b \wedge \dfrac{\dfrac{x = x'}{Sem_E(x,x,y,x')}}{Sem_S(x = x, x, y, x', y')}}{Sem_{s_{max2}} = Sem_S(\texttt{Ite (x < y) (x = y) (x = x)}, x, y, x', y')} \quad (2)$$

Because the tree is finite, the relation $Sem_{s_{max2}}$ can be defined by a recursion-free formula. In particular, we can "symbolically execute" the tree in Equation (2) starting from the leaves and working toward the root. At each step, a semantic relation in the succedent of an inference-rule instance is replaced by its definition and simplified using the properties available in the antecedent. Via this process, we can extract a formula that exactly characterizes $Sem_{s_{max2}}$, which we then substitute for $Sem_{s_{max2}}$ in Equation (1). We thus obtain the following equivalent formula ($\varphi_{SEM_{s_{max2}}}$), which is stated entirely in first-order logic, without any fixed-point operators:

$$\forall x, y, x'. \, (\exists y'. y = y' \wedge ((x < y \wedge x' = y) \vee (x \geq y \wedge x' = x))) \Leftrightarrow (x' = x \vee x' = y) \wedge x \leq x' \wedge y \leq x'$$

In our tool Muse, the quantified satisfiability modulo theories (SMT) solver Z3 [10] proves this formula valid in 0.06 seconds, proving that $s_{max2}$ is a correct solution to this SemGuS problem.

## 2.2 DoubleViaLoop: Partial (CHCs) and Total Correctness ($\mu$CLP)

While CHCs were insufficient for reasoning about the example presented in §2.1, a large class of interesting SemGuS verification problems can be solved as a CHC satisfiability problem—i.e., SemGuS problems whose specifications encode partial correctness. Consider the SemGuS problem given in Figure 2, which requires synthesizing an imperative program (this time potentially containing a loop) that is partially correct. In particular, if the variables x and y start with the values $x$ and $y$, respectively, and the program terminates, then it must set the value $y'$ of variable y to $2x$. The grammar $G_{loop}$ is restricted so that assignments can only increment and decrement variables (Figure 2a)—i.e., a correct program for the task must contain a loop. The semantics $Sem_{loop}$ of this language (Figure 2d) is defined similarly to the one in our previous example. The key distinction is how the second CHC, which defines the (big-step) semantics of loops, is not structurally decreasing—i.e., the loop $l$ appears again in a semantic relation in the premise of the CHC. Similar to the previous example, we can prove $s_{loop}$ correct by proving validity of the query $Q_{loop} \triangleq Sem_{loop}^{LFP} \models \varphi_{loop}[f_{loop} \mapsto s_{loop}]$. Because $\varphi_{loop}$ contains only negative occurrences of the semantic relations, one can solve the query using the typical approach for CHCs. In our tool Muse, we use the CHC solver Spacer [30] to solve the query in 0.2s, proving that $s_{loop}$ is a valid solution.

*Total Correctness.* Consider the following alternative specification, $\varphi_{loop}^{tot}$, which states a form of total correctness—if $x$ is positive and $y'$ is twice $x$, then starting in the state where x is $x$ and y is 0 the candidate program *must terminate* in a state where x takes value 0 and y takes value $y'$:

$$\forall x, y'. \, 0 \leq x \wedge 2x = y' \Rightarrow Sem_L(f_{loop}, x, 0, 0, y')$$

The above formula ensures termination by having $Sem_L$ appear positively in the specification—i.e., for a program $t$ to satisfy the specification, if $0 \leq x \wedge 2x = y'$, then $Sem_L(t, x, 0, 0, y')$ must be inhabited (i.e., because $Sem_L$ is the least solution to the set of semantic rules in Figure 2d, the proof-tree of $Sem_L(t, x, 0, 0, y')$ must have finite height for all values $x$ and $y'$ can take). The reason why this specification ensures total correctness is because the semantic rules defining $Sem_L$ are deterministic. If the semantics of $Sem_L$ were non-deterministic, this specification would only ensure that there is *at least one* terminating execution, not that *all* executions are terminating.

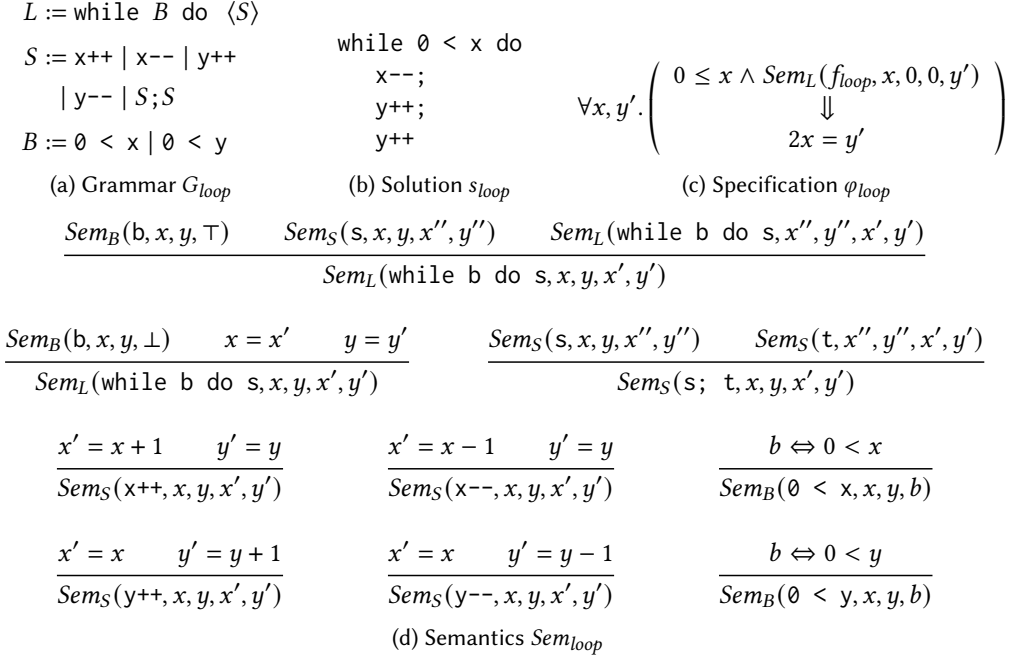

Fig. 2. The four components of the specification of "multiply by 2" in the language $G_{loop}$.

Similar to the example in §2.1, the specification contains a positive occurrence of a semantic relation and thus cannot be solved via reduction to CHC satisfiability. However, unlike the example in §2.1, the program $s_{loop}$'s semantics does not have a finite derivation tree (i.e., the semantics of loops are necessarily recursive). To prove the query $Q_{loop}^{tot} \triangleq Sem_{loop}^{LFP} \models \varphi_{loop}^{tot}[f_{loop} \to s_{loop}]$, we turn to a different approach. Rather than CHCs, we consider a more expressive logic, $\mu$CLP [40].

The $\mu$CLP calculus is a fixed-point logic that generalizes CHCs. Specifically, we can translate the CHCs in $Sem_{loop}$ into *least fixed-point* equations within the $\mu$CLP calculus (whose one and only solution is exactly $Sem_{loop}^{LFP}$). The process to translate from CHCs to fixed-point equations in $\mu$CLP is straightforward. A CHC of the form $\forall \overline{x}_0, \ldots, \overline{x}_n. H(\overline{x}_0) \leftarrow R_1(\overline{x}_1) \wedge R_n(\overline{x}_n) \wedge \varphi$ is translated to the least fixed-point equation $H(\overline{x}_0) =_\mu \exists \overline{x}_1, \ldots, \overline{x}_n. R_1(\overline{x}_1) \wedge R_n(\overline{x}_n) \wedge \varphi$. For a set of CHCs, the translation is applied to each CHC, then equations with identical left-hand sides are combined disjunctively. Thanks to the first-class support for least fixed-points in the $\mu$CLP calculus, we can exactly translate $Q_{loop}^{tot}$—and all SemGuS verification queries—into an equivalent $\mu$CLP query. In our tool MUSE, we perform this translation and solve the produced $\mu$CLP query using MuVal ([40]) in 1.6s, proving that $s_{loop}$ is totally correct.

## 2.3 Beyond SemGuS

§2.2 showed that for every SemGuS problem, one can verify the correctness of a candidate solution using a $\mu$CLP solver. This connection raises a natural question in the opposite direction: *Is there an extension of SemGuS that can express a class of interesting synthesis problems whose semantics and properties of interest cannot currently be encoded in SemGuS, but whose verification conditions are expressible within $\mu$CLP?* In this paper, we answer the question affirmatively and propose SemGuS$^\mu$,

Verifying Solutions to Semantics-Guided Synthesis Problems 7

$$S := \text{repeat } S \mid \text{stay} \mid L \mid R$$
$$\mid L; S \mid R; S$$

```
repeat
  R; R; R; R; R;
  L; L; L; L; L
```

$$\neg \overline{Buchi}(strat, 0, 0)$$

(a) Robot Strategy  (b) Solution *strat*  (c) Specification

$$\overline{Buchi}(strat, x, y) \leftarrow x \neq y \lor (\exists y'. \, 0 \leq y' \leq 5 \land \neg Reach(strat, x, y'))$$

$$Reach(strat, x, y) \leftarrow \neg \overline{Buchi}(strat, x, y) \lor (\exists x', strat'. \, Move(strat, x, x', strat') \land reach(strat', x', y))$$

$$Move(\text{repeat } s, x, x', strat') \leftarrow Move(s; \text{repeat } s, x, x', strat')$$

$$Move(L; s, x, x', strat') \leftarrow strat' = s \land x' = x - 1$$

$$Move(R; s, x, x', strat') \leftarrow strat's \land x' = x + 1$$

(d) Semantics $Sem_{Buchi}$.

Fig. 3. An example of a SemGuS$^\mu$ problem encoding a Büchi game (a kind of reactive synthesis problem). The Büchi game requires the player (a robot) to follow a given strategy to forever reach a sequence of moving targets. The set of allowable strategies is displayed in (a). The robot can move left or right (possibly forever using repeat). In (b) a solution satisfying the Büchi game is displayed. Following *strat* the robot will repeatedly patrol right and left five paces. The specification in (c), requires the robot to reach the moving target forever, when starting at the origin. In (d) we express the rules of the Büchi game as well as the semantics of the productions used to define the solution.

a relatively minor extension of SemGuS such that, in a sense for which we provide a formal proof in Theorem 4.7, SemGuS$^\mu$ *captures exactly* every synthesis problem for which solutions can be verified using $\mu$CLP.

We illustrate this extension with the SemGuS$^\mu$ synthesis problem shown in Figure 3, which requires synthesizing a strategy for a robot to reach a series of targets infinitely often. These types of synthesis problems are often referred to as reactive synthesis problems. For simplicity, we consider a world in which the robot and target's positions are represented by integers with the targets appearing within a bounded region (e.g., between 0 and 5). Once the robot reaches a target, an adversary picks the location of the next target and the game continues.

In Figure 3b, we depict a strategy, *strat*, for the robot. Intuitively, the strategy represents the robot patrolling left and right within a bounded region (i.e., *strat* instructs the robot to move five units right, then five units left, and repeat). To verify that the strategy *strat* results in the robot winning the game in Figure 3, we generate the following verification query:

$$Sem_{Buchi}^{FP} \models \neg \overline{Buchi}(strat, 0, 0) \quad (3)$$

We call the semantic relation $\overline{Buchi}$ because its dual (along with *Reach*) defines a Büchi game. In general, Büchi games are played between two players—the first player tries to reach a goal infinitely often, while the second player tries to thwart the first player. Intuitively, the right-hand side of the verification query encodes when the robot (using the strategy *strat*) wins the Büchi game; the left-hand side of the query ($Sem_{Buchi}^{FP}$) defines the rules of the game. Intuitively, *Reach* encodes that the robot must eventually satisfy the Büchi condition (i.e., denoted by $\neg \overline{Buchi}(strat, x, y)$). The Büchi condition is represented implicitly as the negation of its dual $\overline{Buchi}$. The Büchi condition states that the robot must have reached the target, and then an adversary gets to chose a new target and the game repeats, forever.

To solve the verification query in Equation (3), we must compute the fixed-point of $Sem_{Buchi}$. Unlike the previous examples, the semantics $Sem_{Buchi}$ is not defined using CHCs—most notably due to the negative occurrences of $\overline{Buchi}$ and *Reach* within the premise of the semantic rules. In fact, $Sem_{Buchi}$ does not even define a least fixed-point (due to the negative occurrences). However, such alternating least and greatest fixed-points can be defined in $\mu$CLP. Our tool Muse dispatches this query to the μCLP validity solver MuVal [40], which proves the above query valid in 21s, thereby proving that the strategy *strat* is a valid solution to the SemGuS$^\mu$ synthesis problem in Figure 3.

## 3 SEMGUS AND SEMGUS$^\mu$

This section reviews the SemGuS framework [28] and describes the more expressive framework SemGuS$^\mu$ we propose. A SemGuS synthesis problem is defined in three parts: a grammar defining the syntax of the language over which programs are to be synthesized (§3.1), a set of logical formulas defining the semantics of programs in the language (§3.2), and a specification defining the properties the synthesized program should exhibit (§3.3).

### 3.1 Syntax as Regular Tree Grammars

The syntax of a programming language is defined as a typed regular tree grammar (RTG). A *ranked alphabet* is a tuple $\langle \Sigma, rk_\Sigma \rangle$ consisting of a finite set of symbols ($\Sigma$) and a function $rk_\Sigma : \Sigma \to \mathbb{N}$ that associates every symbol with a rank. For any $n \geq 0$, $\Sigma^n \subseteq \Sigma$ denotes the set of symbols of rank $n$. The set of all *(ranked) Trees* over $\Sigma$ is denoted by $T_\Sigma$. Specifically, $T_\Sigma$ is the least set such that $\Sigma^0 \subseteq T_\Sigma$ and if $\sigma^k \in \Sigma^k$ and $t_1, \ldots, t_k \in T_\Sigma$, then $\sigma^k(t_1, \ldots, t_k) \in T_\Sigma$. In the remainder, we assume a fixed ranked alphabet $\langle \Sigma, rk_\Sigma \rangle$.

*Definition 3.1 (Regular Tree Grammar).* A *typed Regular Tree Grammar* (RTG) is a tuple $G = \langle N, \Sigma, T, a, \delta \rangle$, where $N$ is a finite set of non-terminal symbols of rank 0; $\Sigma$ is a ranked alphabet; $T = \{\tau_0, \ldots, \tau_k\}$ is a finite set of types; $a$ is a type assignment assigning each non-terminal to a type and each symbol of rank $i$ to a tuple of of types $\langle \tau_0, \ldots, \tau_i \rangle \in T^{i+1}$; and $\delta$ a finite set of productions of the form $A_0 \to \sigma^i(A_1, \ldots, A_i)$ such that for all $0 \leq j \leq i$, $A_j \in N$ is a non-terminal and if $a_{\sigma^i} = \langle \tau_0, \ldots, \tau_i \rangle$ then $a_{A_j} = \tau_j$.

Given a tree $t \in T_{\Sigma \cup N}$, one may apply the production rule $r = A \to \beta \in \delta$ to $t$ to produce a tree $t'$ by replacing the leftmost occurrence of $A$ in $t$ with $\beta$. A tree $t \in T_\Sigma$ is generated by the grammar $G$ ($t \in L(A)$) when $t$ is the result of applying some sequence of production rules $r_0, \ldots, r_n \in \delta^n$ to a non-terminal $A \in N$.

*Example 3.2 (RTG).* For example, the syntax of programs considered in Figure 2a represents a regular tree grammar. It consists of the nonterminals $L$, $S$, and $B$; ranked symbols while$^2$, x++$^0$, x--$^0$, y++$^0$, y--$^0$, seq$^2$, 0 < x$^0$, and 0 < y$^0$, and productions $L \to$ while$(B,S)$, $S \to$ x++, $S \to$ x--, $S \to$ y++, $S \to$ y--, $S \to$ seq$(S,S)$, $B \to$ 0 < x, and $B \to$ 0 < y. In the examples in the paper, we often drop the ranks of symbols and use infix notation to enhance readability.[1]

### 3.2 Semantics via Logical Relations and Fixed-point Logics

We begin by reviewing some necessary details of the fragments of first-order logic we use in this paper. Given a (possibly multi-sorted) first-order theory $\mathcal{T}$ over a signature $\Sigma$, the syntax of formulas and terms are given by the following grammar:

$$\varphi ::= X(t_1, \ldots, t_{rk_\Sigma(X)}) \mid p(t_1, \ldots, t_{rk_\Sigma(p)}) \mid \neg \varphi_1 \mid \varphi_1 \wedge \varphi_2 \mid \forall x : s.\varphi_1$$
$$t ::= x \mid f(t_1, \ldots, t_{rk_\Sigma(f)})$$

where $x$ and $X$ are term and predicate variables, respectively; $f$ and $p$ are function and predicate symbols of $\Sigma$; and $s$ is a sort of $\Sigma$. Disjunction, implication, existential quantification, etc. are omitted from the syntax and may be defined as expected (e.g., $\varphi \lor \psi \triangleq \neg(\neg\varphi \land \neg\psi)$). We will use $\varphi$ and $\psi$ to refer to possibly quantified formulas, and $F$ and $G$ to refer to quantifier-free formulas. We use $FV(\varphi)$ and $FV(t)$ to denote the free variables of a formula and term, respectively. Given a formula $\varphi$, variable $x$, and term $t$, we use $\varphi[x \mapsto t]$ to denote $\varphi$ with every free occurrence of $x$ replaced with $t$. Additionally, for a set of variables $V$, we use $\varphi[V \mapsto c_x]$ to represent replacing every free occurrence of each $x \in V$ with a constant $c_x$.

---
[1]The grammars used in the paper are referred to at various places as "grammars" or "regular-tree grammars" (Defn. 3.1). The trees/terms in the language of a grammar would be represented using algebraic data types. In the logics used in the paper (CHCs, co-CHCs, and μCLP), we implicitly assume that one can use values in the algebraic data type to express tree-valued constants.

A constrained Horn clause (CHC) is a formula over some background theory of the form:

$$\forall \bar{x}_0, \ldots, \bar{x}_n. X_0(\bar{x}_0) \leftarrow X_1(\bar{x}_1) \land \cdots \land X_n(\bar{x}_n) \land F(\bar{x}_0, \ldots, \bar{x}_n), \tag{4}$$

where each $\bar{x}_i$ is a sequence of term variables, $X_i$ is a predicate variable, and $F$ is a constraint over the variables in each predicate. In the remainder of the paper, we abuse notation and allow arbitrary first-order terms to appear as arguments to each $X_i$.

In the SemGuS framework originally defined by Kim et al. [28], the semantics of programs in the language defined by the regular tree grammar is provided by defining a logical relation and using CHCs over some theory, to define the elements of the relation by giving rules for each of the productions of the grammar. As discussed in §2.3, in our work, we represent the input semantics in a logic more expressive than CHCs—in particular, relations are ordered, and their definitions can include quantified variables and conjunctive and disjunctive combinations of constraints and positive and negative occurrences of semantic relations.

*Definition 3.3 (SemGuS$^\mu$ semantics).* Given a first-order theory $\mathcal{T}$ and regular tree grammar $G = \langle N, \Sigma, T, a, \delta \rangle$ a semantics for $G$ is a tuple $\langle SEM, <_{SEM}, [\![\cdot]\!] \rangle$ where $SEM$ maps each non-terminal $A \in N$ to a non-empty finite set of uninterpreted relations ($SEM_A = \{Sem_A^1, \ldots, Sem_A^n\}$), $<_{SEM}$ is a total ordering over all semantic relations, and $[\![\cdot]\!]$ maps each production rule $A_0 \to \sigma^i(A_1, \ldots, A_i)$ of type $(\tau_0, \ldots, \tau_i)$ and semantic relation $Sem_{A_0}^j \in SEM_{A_0}$ to a formula of the form $Sem_{A_0}^j(t_{A_0}, \Gamma^{0,j}, \Upsilon^0) \leftarrow \varphi$ such that:

- $\varphi$ is a (possibly quantified) $\mathcal{T}$ formula,
- $t_{A_0}$ is a variable representing elements of $L(A_0)$, $\Upsilon^0$ is a variable of type $\tau_0$, and $\Gamma^{0,j}$ are variables representing state,
- $\varphi$'s free variables belong to $\Gamma^{0,j}$, $\Upsilon^0$, or $\{t_{A_0}\}$,
- $Sem_{A_0}^j$ does not appear negatively within $\varphi$ (either directly within $\varphi$ or indirectly within the definition of the semantic relations appearing in $\varphi$), and
- For each $Sem_{A_k}^l(t_{A_k}, \Gamma^{k,l}, \Upsilon^k)$ appearing in $\varphi$:
  - $0 \leq k \leq i$ and $Sem_{A_k}^l \in SEM_{A_k}$ and
  - $t_{A_k}$, $\Upsilon^k$, and $\Gamma^{k,l}$ are defined analogously to $t_{A_0}$, $\Upsilon^0$, $\Gamma^{0,j}$.

*Example 3.4.* Consider the semantics $Sem_{loop} = \langle SEM, <_{SEM}, [\![\cdot]\!] \rangle$ in Figure 2d. Each non-terminal is mapped to a single semantic relation (i.e., $SEM_L = \{Sem_L\}$, $SEM_S = \{Sem_S\}$, and $SEM_B = \{Sem_B\}$). In this instance, order does not matter, and we can assume an arbitrary order, such as $Sem_L <_{SEM} Sem_S <_{SEM} Sem_B$. The semantic function $[\![\cdot]\!]$ maps each semantic relation $Sem_A$ and production rule $A \to \sigma^i(A_1, \ldots, A_i)$ to the semantic relation whose head is of the form $Sem_A(\sigma^i(t_1, \ldots, t_n), \Gamma, \Upsilon)$. For example, $[\![0 < x]\!]_{Sem_B}$ is the rule $Sem_B(0 < x, x, y, b) \leftarrow 0 < x$.

Our semantics generalizes the semantic rules considered by Kim et al. [28] in three ways:

(1) It allows each nonterminal to be associated with *multiple* semantic relations—e.g., to describe the multiple relations appearing in the example from Figure 3.
(2) The rules defining the semantic relations are expressed in a fragment of first-order logic that goes beyond CHCs—e.g., to describe the rules that define $Reach_T$ used in Figure 3.

(3) Each semantic relation is ordered to ensure that the semantic rules correspond to a unique fixed-point describing the semantics of the language.

If we restrict our semantic definition to have a single semantic relation per non-terminal and to rules of the form $Sem_A(t_A, \Gamma, \Upsilon) \leftarrow \varphi$, where $\varphi$ contains only existential quantification and positive occurrences of semantic relations, then our definition is equivalent to the semantics considered in SemGuS [28]. Note that, while we allow only one rule per production per semantic relation, we do allow for the disjunction of semantic relations within the premise of a rule, thereby recovering equivalent expressiveness to allowing multiple rules per production rule. In the remainder of this paper, to enhance readability we omit variable types from semantic relations when appropriate (i.e., we write $Sem_A(t, \overline{x})$ instead of $Sem_A(t, \Gamma, \Upsilon)$). The robot-reachability synthesis problem considered in Figure 3 cannot be encoded in SemGuS, but can be encoded in SemGuS$^\mu$.

## 3.3 Specifications and SemGuS$^\mu$ Problems

Now that we have a way to define the syntax and semantics of the programming language over which we are trying to synthesize programs, all that is missing to define a SemGuS$^\mu$ problem is the specification we want the synthesized program to satisfy.

*Definition 3.5 (SemGuS$^\mu$ problem, solution, validity, realizable).* A SemGuS$^\mu$ problem is a tuple $\mathcal{P} = \langle G = \langle N, \Sigma, T, a, \delta \rangle, \langle SEM, <_{SEM}, [\![\cdot]\!] \rangle, F, \varphi \rangle$, where

- $G$ is a regular tree grammar.
- $\langle SEM, <_{SEM}, [\![\cdot]\!] \rangle$ is a semantics for $G$.
- $F$ is a finite set of functions we want to synthesize—pairs of the form $\langle f, A \rangle$ where $f$ is a variable representing a procedure that we want to synthesize, and $A \in N$ is the root nonterminal from which $f$ is to be derived—i.e., the solution for $f$ must be a tree $t \in L(A)$.
- $\varphi$ a specification in the theory $\mathcal{T}$ such that
  - The free variables of $\varphi$ must be functions to synthesize, $FV(\varphi) \subseteq \{f : \langle f, A \rangle \in F\}$ and
  - For any $\langle f, A \rangle \in F$, $f$ appears only in atoms of the form $Sem_A^i(f, \overline{x})$ where $Sem_A^i \in SEM_A$.

For a semantics $\langle SEM, <_{SEM}, [\![\cdot]\!] \rangle$, an *interpretation* $\rho$ is a function that maps each semantic relation $Sem_A^i(t, \overline{x}) \in SEM$ to a formula whose free variables are $\overline{x} \cup \{t\}$. The interpretation $SEM^{LFP}$ is the interpretation that maps each semantic relation to its least solution.

A *solution* to the SemGuS$^\mu$ problem $\mathcal{P}$ is a function $S$ that maps each $\langle f, A \rangle \in F$ to a tree $t \in L(A)$. The solution $S$ is *valid* when $SEM^{LFP} \models \varphi[\langle f, A \rangle \in F.f \mapsto S(f)]$. Note that the values $S(f)$ being substituted into the formula are program-valued constants represented as terms in the algebraic data type for $G$. Moreover, by the last case of Definition 3.5, each occurrence of $f$ in $\varphi$ is in an atom of the form $Sem_A^i(f, \overline{x})$ where $Sem_A^i \in SEM_A$. Consequently, in the resulting formula, each such program-valued constant will be interpreted according to the least fixed-point of a semantic relation of an appropriate kind. Note that, while $SEM^{LFP}$ appears to use only least fixed-points, because we allow semantic relations to appear negated within the premise of a semantic rule, $SEM^{LFP}$ represents arbitrarily nested greatest and least fixed-points whose top-level fixed-point is a least fixed-point. We say that $\mathcal{P}$ is *realizable* if there exists a valid solution to $\mathcal{P}$.

In SemGuS$^\mu$ we allow multiple (mutually recursive) semantic relations to afford a SemGuS$^\mu$ user flexibility when defining a semantics. For example, consider Figure 3, which uses this flexibility to define the semantics of a Büchi game. Without multiple semantic relations, we could not define both $\overline{Sem_{Büchi}}$ and $Sem_{Reach}$. Furthermore, because we allow semantic relations to appear negatively within the body of semantic rules, the order of semantic rules can affect the interpretation of the rules! For example, consider the mutually recursively defined semantic relations $Sem_A$ and $Sem_B$, whose semantics follows a similar pattern to the semantic relations appearing in Figure 3d:

$$Sem_A(x) \leftarrow Sem_A(x) \vee \neg Sem_B(x)$$
$$Sem_B(x) \leftarrow Sem_B(x) \vee \neg Sem_A(x)$$

The least fixed-point of $Sem_A$ requires first computing the greatest fixed-point of $\overline{Sem_B}$, and similarly the least fixed-point of $Sem_B$ requires first computing the greatest fixed-point of $\overline{Sem_A}$. To ensure that the semantic rules correspond to a unique fixed-point, we fix the order of evaluation of the semantic relations. For complete details we refer the reader to [40]. If $Sem_A <_{SEM} Sem_B$, then $Sem_A$ is defined by the fixed-point $Sem_A(x) =_\mu Sem_A(x) \vee \overline{Sem_B}(x); \overline{Sem_B}(x) =_\nu \overline{Sem_B}(x) \wedge Sem_A(x)$ (which evaluates $Sem_A$ to $\emptyset$). Otherwise, $Sem_A$ is defined by $\overline{Sem_B}(x) =_\nu \overline{Sem_B}(x) \wedge Sem_A(x); Sem_A(x) =_\mu Sem_A(x) \vee \overline{Sem_B}(x)$ (which evaluates $Sem_A$ to $\mathbb{Z}$).

## 4 VERIFYING CANDIDATE PROGRAMS

This section formalizes the three methods used in §2 to verify that a program is a valid solution to a SemGuS problem. Each technique encodes when the program is a valid solution to the SemGuS problem in a fragment of first-order logic (with fixed-points). We describe each of the three encodings, and characterize the kinds of SemGuS verification problems on which they can be applied (§§ 4.1 to 4.3). Additionally, we prove that the SemGuS$^\mu$ framework described in §3 can be used to define verification problems that require the full capabilities of $\mu$CLP. We now describe each encoding in turn. In the remainder of this section, we consider a fixed SemGuS$^\mu$ problem $\mathcal{P} = \langle G = \langle N, \Sigma, S, T, a, \delta \rangle, \langle SEM, [\![\cdot]\!] \rangle, F, \varphi \rangle$ and candidate solution $P$.

### 4.1 Encoding Nonrecursive SemGuS$^\mu$ Verification Problems with Quantified SMT

In §2.1, we were able to produce a first-order-logic formula that is free of any semantic relations and is satisfiable exactly when $\varphi_{max2}$ is a valid solution to the SemGuS problem displayed in Figure 1. We could obtain such a formula because the derivation tree of the semantics of $\varphi_{max2}$ is finite. To formalize this intuition, we define two auxiliary notions: when a semantic relation is non-recursive on tree $t$, and when a semantic relation is a $t$-ancestor of another semantic relation—i.e., when the semantic relations are not recursive on the program term.

*Definition 4.1 (t-ancestor, non-recursive on t).* Let $A \in N$ be any non-terminal, $t \in L(A)$ be a tree of the form $t = \sigma^i(t_1, \ldots, t_i)$ for some production rule $A \to \sigma^i(A_1, \ldots, A_i)$, semantic relation $Sem_A \in SEM_A$ of $A$, and $[\![A \to \sigma^i(A_1, \ldots, A_i)]\!]_{Sem_A} = Sem_A(t, \overline{x}) \leftarrow \varphi$.

We say that a semantic relation $Sem'_A \in SEM_A$ is a *t-ancestor* of $Sem_A$ if and only if (i) $Sem'_A(t, \overline{x})$ appears in the antecedent $\varphi$ for some $\overline{x}$, or (ii) there is some symbol $Sem''_A(t, \overline{x})$ that appears in $\varphi$ and $Sem'_A$ is a $t$-ancestor of $Sem''_A$.

We say that $Sem_A$ is *non-recursive on t* if (i) $Sem_A$ is not a $t$-ancestor of itself, and (ii) for each $Sem_{A_j}(t_j, \overline{x})$ appearing in $\varphi$, $Sem_{A_j}$ is non-recursive on $t_j$. If $Sem_A$ is non-recursive on $t$ then for any arguments $\overline{x}$, the derivation tree of $Sem_A(t, \overline{x})$ has finite height.

*Definition 4.2 (Formula of).* Given a non-terminal $A \in N$, a semantic relation $Sem_A \in SEM_A$, and a production $A \to \sigma^i(A_1, \ldots, A_n) \in \delta$, if $[\![A \to \sigma^i(A_1, \ldots, A_n)]\!]_{Sem_A}$ is of the form $Sem_A(t, \overline{x}) \leftarrow \varphi$, then the formula of $Sem_A(t', \overline{x}')$ (denoted by $\varphi$-of$(Sem_A(t', \overline{x}'))$) is $\varphi[t \mapsto t', \overline{x} \mapsto \overline{x}']$, which replaces the formal arguments of $Sem_A$ with the actual arguments of the application.

We now turn to defining the procedure SMT-FORMULA-OF that encodes that a solution $P$ is valid for a SemGuS$^\mu$ problem (where the semantics is non-recursive on $P$) into first-order logic without fixed-points (i.e., quantified SMT formulas). SMT-FORMULA-OF repeatedly replaces every occurrence of a semantic relation with the premise of the rule that defines it.

```
1  Procedure SMT-FORMULA-OF(𝒫 = ⟨G, ⟨SEM, ⟦·⟧⟩, F, φ⟩, S)
2      rules ← ⊤                                       // empty set of rules to begin with
3      φ ← φ[⟨f, A⟩ ∈ F.f ↦ S(f)]                      // substitute solution into specification
4      foreach $Sem_A(t, \overline{x})$ appearing in φ do
5          ψ ← φ-of($Sem_A(t, \overline{x})$)          // repeatedly replace semantic relations with their def.
6          while $Sem_{A'}(t', \overline{x}')$ appears in ψ do
7          ⌊ ψ ← ψ[$Sem_{A'}(t', \overline{x}')$ ↦ φ-of($Sem_{A'}(t', \overline{x}')$)]
8          rules ← rules ∧ ($Sem_A(t, \overline{x})$ ⇔ ψ)    // update rules to add definition for $Sem_A$
9      return ⟨rules, φ⟩                               // $rules^{LFP}$ ⊨ φ if and only if S is valid solution to 𝒫
```

Applying SMT-FORMULA-OF to the verification problem in §2.1 yields the formula in Equation (1). The following theorem states under which conditions SMT-FORMULA-OF($\mathcal{P}, P$) returns a formula that is satisfiable if and only if $P$ is a valid solution to $\mathcal{P}$.

THEOREM 4.3 (SMT-FORMULA-OF IS SOUND). *For any SemGuS$^\mu$ problem $\mathcal{P} = \langle G = \langle N, \Sigma, S, T, a, \delta \rangle, \langle SEM, \llbracket \cdot \rrbracket \rangle, F, \varphi \rangle$ and solution $P$ of $\mathcal{P}$, if $Sem_A$ is non-recursive on $P(f)$ for each occurrence of $Sem_A(f, \overline{x})$ within the specification $\varphi$, then SMT-FORMULA-OF($\mathcal{P}, P$) is valid if and only if $P$ is a valid solution of $\mathcal{P}$.*

### 4.2 Encoding CHC-like SemGuS$^\mu$ Verification Problems with CHCs

In §2.2, we saw how to encode the SemGuS verification problem from Figure 2 into the CHC fragment of first-order logic when using the specification $\varphi_{loop}$ from Figure 2c. In this section, we formalize when and how a SemGuS verification problem may be encoded with CHCs.

To encode the verification problem a CHC decision problem, we require the semantics of the solution to be equivalent to a set of CHCs (i.e., formulas of the form described in Equation (4)). For CHCs the decision problem of interest is "given a set of CHCs and a query formula of the form $\forall \overline{x}.R(x) \Rightarrow \varphi$, determine if the query is derivable from the set of CHCs"—or, equivalently, "determine if some interpretation of the uninterpreted relations satisfies all relations and the given query formula" [9]. Furthermore, for the given class of queries, the problem is also equivalent to determining if the least solution to the set of CHCs satisfies the desired query [25]. We use this final notion to formulate our verification procedure CHC-OF.

*Definition 4.4 (CHC-like).* Let $A \in N$ be any non-terminal, $t \in L(A)$ be a tree of the form $t = \sigma^i(t_1, \ldots, t_i)$ for some production rule $A \to \sigma^i(A_1, \ldots, A_i)$, $Sem_A \in SEM_A$ a semantic relation of $A$, and $\llbracket A \to \sigma^i(A_1, \ldots, A_i) \rrbracket_{Sem_A} = Sem_A(t, \overline{x}) \leftarrow \varphi$.

We say the rules defining $Sem_A$ are *CHC-like for t* if and only if (i) $\varphi$ has no negative occurrences of a semantic relation, (ii) $\varphi$ contains no universal quantifiers, and (iii) for every $Sem_{A_j}(t_j, \overline{x}_j)$ appearing in $\varphi$, the rules defining $Sem_{A_j}$ are CHC-like for $t_j$.

For example, the rules defining both $Sem_{max2}$ and $Sem_{loop}$ in Figures 1d and 2d are CHC-like (for any program within their respective grammars), while the rules for $Sem_{Buchi}$ in Figure 3d are not. We now define the procedure CHC-OF, which encodes as a CHC satisfiability problem the property that a solution $P$ is valid for a SemGuS$^\mu$ problem (where the semantics is CHC-like for $P$). We first define an auxiliary function *rules-of* that, given a semantic relation $Sem_A$ and tree $t \in L(A)$, returns a set of CHCs logically equivalent to the semantic rule defining $Sem_A$ for the root production of $t$. Let $Sem_A(t, \overline{x}) \leftarrow \varphi = \llbracket A \to \sigma^i(A_1, \ldots, A_i) \rrbracket_{Sem_A}$ and let $A \to \sigma^i(A_1, \ldots, A_i)$ denote the root production of $t$. Under our assumptions (that $Sem_A$ is CHC-like on $t$), we may assume that $\varphi$ is a formula that may contain existential quantification and arbitrary disjunction and conjunction of positive occurrences of semantic relations. Without loss of generality, we assume

```
1  Procedure CHC-OF(⟨G = ⟨N, Σ, S, T, a, δ⟩, ⟨SEM, ⟦·⟧⟩, F, φ⟩, S)
2      rules ← ⊤ ;
3      ψ ← φ[⟨f, A⟩ ∈ F.f ↦ S(f)];
4      Q ← {⟨Sem_A, t⟩ : Sem_A(t, x̄) appears in ψ} ;    // Queue of relations that need defining
5      while Q ≠ ∅ do
6          ⟨Sem_A, t'⟩ ← pick Q ;
7          rules' ← rules-of (Sem_A, t') ;              // Definition of Sem_A for t' as a set of CHCs
8          Q ← Q ∪ {⟨Sem_{A_j}, t_j⟩ : Sem_{A_j}(t_j, x̄_j) appears negatively in rule[t ↦ t']
9                             for some rule in rules'} ;
10         rules ← rules ∧ ⋀ rules' ;                    // Add rules defining Sem_A for t' to rules
11     return ⟨rules, ψ⟩ ;                              // rules^{LFP} ⊨ ψ if and only if S is a valid solution to 𝒫
```

that any existentially bound variable in $\varphi$ is uniquely named and does not appear in $\overline{x}$ (otherwise, it can be renamed). Let $F$ be the quantifier-free formula obtained by erasing all existential quantifiers from $\varphi$. Then $dnf(\varphi)$ is the set of disjuncts of the dnf of $F$. Finally, define $\textit{rules-of}(\textit{SemA}, t)$ to be the set $\{\textit{Sem}_A(t, \overline{x}) \leftarrow \psi : \psi \in dnf(\varphi)\}$. The CHC-OF procedure produces a set of CHCs that captures the semantics of each candidate program $t$ by effectively performing a breadth-first search on each of the semantic relations that define the semantics of each sub-program of $t$. In Theorem 4.5, we state conditions under which CHC-OF is sound.

While one might expect the CHC-encodable fragment of SemGuS$^\mu$ to subsume the SMT encodable fragment of SemGuS$^\mu$—e.g., by encoding the output of SMT-FORMULA-OF into an equi-satisfiable set of CHCs—the two fragments are incomparable. Rather than performing complex encodings, both SMT-FORMULA-OF and CHC-OF consider fragments of SemGuS$^\mu$ that can be naturally encoded into SMT and CHCs, respectively. Specifically, SMT-FORMULA-OF only considers programs whose semantics recurses on structurally smaller programs (even if semantic relations appear positively or negatively in the specification or the semantic definitions). In contrast, CHC-OF only considers semantics that can be automatically translated into an equivalent set of CHCs (i.e., by restricting semantic rules to only include existential quantifiers and positive occurrences of other semantic relations, even those that recurse on non-structurally decreasing terms). For example, SMT-FORMULA-OF can encode the SemGuS problem found in fig. 1, because it involves a non-recursive program but could not encode the SemGuS problem found in fig. 2 because it requires reasoning about a recursive program. In contrast, CHC-OF cannot encode the example in fig. 1, because the specification includes both positive and negative occurrences of a semantic relation, whereas CHC-OF can encode the example in fig. 2 because the semantics is represented as CHCs and the specification only contains a negative occurrence of a semantic relation.

THEOREM 4.5 (CHC-OF IS SOUND.). *For any SemGuS$^\mu$ problem $\mathcal{P} = \langle G = \langle N, \Sigma, T, a, \delta \rangle, \langle \textit{SEM}, <_{\textit{SEM}}, \llbracket \cdot \rrbracket \rangle, F, \varphi \rangle$ and solution $P$ of $\mathcal{P}$, if for each occurrence of $\textit{Sem}_A(f, \overline{x})$ within the specification $\varphi$, it appears negatively and the rules defining $\textit{Sem}_A$ are CHC-like for $P(f)$, then the query returned by CHC-OF($\mathcal{P}, P$) is valid if and only if $P$ is a valid solution of $\mathcal{P}$.*

Conversely, in Theorem 4.6, we prove that, in general, a more expressive fragment of first-order logic with fixed-points is required to verify solutions of an arbitrary SemGuS problems. In particular, we need a fragment of the $\mu$CLP calculus that only uses least fixed-point equations.

THEOREM 4.6 (VERIFICATION OF SEMGUS IS NOT REDUCIBLE TO CHC SATISFIABILITY). *There exists a program $t$ and SemGuS problem $Sy$ such that verifying $t$ satisfies $Sy$ cannot be reduced to satisfiability of Constrained Horn Clauses.*

14                  Charlie Murphy, Keith Johnson, Thomas Reps, and Loris D'Antoni

```
1  Procedure MUCLP-OF(𝒫 = ⟨G, ⟨SEM, <_SEM, ⟦·⟧⟩, F, φ⟩, S)
2     rules ← ⊤ ;
3     ψ ← Norm(φ[⟨f, A⟩ ∈ F.f ↦ S(f)]) ;
4     Q ← {⟨Sem_A, t, μ⟩ : Sem_A(t, x̄) appears in ψ} ∪
5         {⟨Sem_A, t, ν⟩ : Sem_A(t, x̄) appears in ψ} ;
6     while Q ≠ ∅ do
7        ⟨Sem_A, t', fix⟩ ← pick Q ;
8        rule ← head-of(Sem_A, t') =_μ Norm(body-of(Sem_A, t')) ;        // Compute as LFP
9        if fix = ν then
10           rule ← dual(rule) ;                                          // Dualize to GFP
11       Q ← Q ∪ {⟨Sem_{A_j}, t_j, μ⟩ : Sem_{A_j}(t_j, x̄_j) appears in body of rule[t ↦ t']};
12       Q ← Q ∪ {⟨Sem_{A_j}, t_j, ν⟩ : Sem_{A_j}(t_j, x̄_j) appears in body of rule[t ↦ t']} ;
13       rules ← rules ∪ {rule} ;
14    rules ← rules sorted by <_SEM ;         // According to the head predicate of each rule.
15    return ⟨rules, ψ⟩ ;                     // rules^FP ⊨ ψ if and only if S is a valid solution to 𝒫
```

### 4.3 Encoding all SemGuS$^\mu$ Verification Problems with $\mu$CLP

In §2.2 and §2.3, we examined two SemGuS$^\mu$ verification problems for which there is no possible encoding as fixed-point-free formulas (i.e., SMT or CHCs). Instead, these problems were encoded into a first-order fixed-point logic, $\mu$CLP, that allows defining both greatest and least fixed-points. Unlike the previous encodings, for any SemGuS$^\mu$ problem $\mathcal{P}$ one can *always* use $\mu$CLP to encode that $P$ is a valid solution to $\mathcal{P}$. A $\mu$CLP formula is a sequence of formulas of the form:

$$X_0(\bar{x}_0) =_{fix_0} \varphi_0 \quad \ldots \quad X_n(\bar{x}_n) =_{fix_n} \varphi_n,$$

where each $X_i$ is an uninterpreted relation, $\bar{x}_i$ is a sequence of term variables, and the $\varphi_i$ are formulas within some background theory whose free variables are $\bar{x}_i$ and which may include positive occurrences of the uninterpreted relations $X_0, \ldots, X_n$. Each $fix_i$ is either $\mu$ or $\nu$ referring to whether or not the equation $X_i(\bar{x}_i) =_{fix_i} \varphi_i$ should represent a least or greatest fixed-point, respectively. We refer the reader to Unno et al. [40] for a detailed formalization of $\mu$CLP.

In the SemGuS$^\mu$ semantics (Definition 3.5), every semantic relation's definition is oriented as a least fixed-point. However, our semantics does allow one to introduce greatest fixed-points by taking the negation of a semantic relation. We now turn to defining MUCLP-OF, which encodes as a $\mu$CLP query the property that a solution $P$ is valid for a SemGuS$^\mu$ problem $\mathcal{P}$. The procedure is similar to CHC-OF, in that it performs a breadth-first search over the semantic relations to produce the resulting $\mu$CLP query. For a formula $\varphi$, we use $Norm(\varphi)$ to denote the formula $\varphi$ wherever a negative occurrence of $Sem_A(t, \overline{x})$ is replaced by $\neg\overline{Sem_A}(t, \overline{x})$ (i.e., so that $\overline{Sem_A}(t, \overline{x})$ appears positively in $Norm(\varphi)$). Next, we define $dual(Sem_A(t, \overline{x}) =_\alpha \varphi)$ to be the dual fixed-point equation $\overline{Sem_A}(t, \overline{x}) =_{\overline{\alpha}} Norm(\neg\varphi)$ where $\overline{\mu} = \nu$ and $\overline{\nu} = \mu$. For all semantic relations $Sem_A$ and programs $t \in L(A)$, let $head\text{-}of(Sem_A, t) \triangleq Sem_A(t, \overline{x})$ and $body\text{-}of(Sem_A, t) \triangleq \varphi$, where $Sem_A(t, \overline{x}) \leftarrow \varphi$ is the semantic relation that defines the semantics of the root production of $t$.

Theorem 4.7 states that MUCLP-OF soundly encodes any SemGuS and SemGuS$^\mu$ problem into a validity query within the $\mu$CLP calculus.

THEOREM 4.7 (MUCLP-OF IS SOUND). *For any SemGuS$^\mu$ problem $\mathcal{P} = \langle G, \langle SEM, <_{SEM}, [\![\cdot]\!] \rangle, F, \varphi\rangle$ and solution $P$ of $\mathcal{P}$, the query returned by MUCLP-OF($\mathcal{P}, P$) is valid iff $P$ is a valid solution of $\mathcal{P}$.*

Theorem 4.8, states that the SemGuS$^\mu$ semantics can express any $\mu$CLP query—i.e., that any $\mu$CLP validity query can be equivalently reduced to a SemGuS$^\mu$ verification problem. Thus, SemGuS$^\mu$ can encode any problem that can be encoded within the $\mu$CLP calculus.

THEOREM 4.8 (SEMGUS$^\mu$ SEMANTICS AND $\mu$CLP ARE EQUALLY EXPRESSIVE). *For every $\mu$CLP query $\langle \varphi, preds \rangle$, there is some SemGuS problem $\mathcal{P}$ and solution $P \in L(G_\mathcal{P})$ such that $\langle \varphi, preds \rangle$ is valid if and only if $P$ is a valid solution to $\mathcal{P}$.*

Conversely, in Theorem 4.9, we state that verification for SemGuS problems does not require the full generality of $\mu$CLP. Specifically, SemGuS verification problems do require a fragment of first-order logic (with fixed-points) beyond both CHCs and coCHCs, but do not require arbitrary alternations of greatest and least fixed-points. As a corollary of Theorems 4.8 and 4.9, SemGuS$^\mu$ is more expressive than SemGuS.

THEOREM 4.9 (SEMGUS AND $\mu$CLP ARE NOT EQUALLY EXPRESSIVE). *Verifying solutions to SemGuS problems can be encoded within a fragment of $\mu$CLP that uses at most one alternation between greatest and least fixed-points.*

## 5 IMPLEMENTATION

We implement our algorithms in a tool, called MUSE, which extends SemGuS to SemGuS$^\mu$. MUSE supports all of the encoding schemes for the three classes of problems discussed in §4.

MUSE is implemented in OCaml, and uses Z3 for SMT formulas [10], Spacer for CHCs [30], and MuVal for $\mu$CLP queries [40]. As part of implementing MUSE, we extended the implementation of MuVal to support algebraic data types, which we use to represent programs in first-order logic.

The remainder of this section describes three optimizations that one may apply to transform the encodings described in §4. The goal of these optimizations is to use knowledge of the SemGuS verification problem to make the resulting optimized queries simpler.

*Reification of Terms in Semantic Relations.* In our verification problems, we have a specific concrete program (or programs) whose semantics we wish to capture using the semantic relations. The goal of the first optimization REIFY is to eliminate program terms from the semantic relations, thus removing the burden of the solver to reason about terms using the theory of algebraic data types.

*Example 5.1.* Consider the program $t \equiv $ x--; y++ from the language $Imp_{loop}$ described in Figure 2. Below we depict the AST of $t$ and show the reified semantics of $Sem_{loop}$ specialized to $t$. Because the program is known *a priori*, the semantics can be reified to remove the AST-valued argument in the different *Sem* relations by introducing a new semantic relation for each node of the AST.

$$Sem_S^{\text{x--; y++}}(x, y, x', y') \leftarrow Sem_S^{\text{x--}}(x, y, x'', y'') \wedge Sem_S^{\text{y++}}(x'', y'', x', y')$$
$$Sem_S^{\text{x--}}(x, y, x', y') \leftarrow x' = x - 1 \wedge y = y'$$
$$Sem_S^{\text{y++}}(x, y, x', y') \leftarrow x' = x \wedge y' = y + 1$$

AST of x--; y++

More formally, the reified semantics introduces a new semantic relation for every sub-tree of the program's AST. Each occurrence of $Sem_{A_j}(t_j, \overline{x}_j)$ is then replaced by $Sem_{A_j}^{t_j}(\overline{x}_j)$.

*Definition 5.2 (Reified Semantics).* Given a non-terminal $A$, a program $t \in L(A)$, a semantic relation $Sem_A$, and a set of semantic rules *rules* defining the semantics of $Sem_A$, the *semantics of $Sem_A$ reified to $t$* is a pair REIFY(*rules*, $Sem_A, t$) = $\langle SEM^{reify}, rules^{reify} \rangle$ such that $SEM^{reify}$ and $rules^{reify}$ are the least solution to the following rules:

(1) $Sem_A^t$ is a reified semantic relation ($Sem_A^t \in SEM^{reify}$),

(2) if $Sem_A^{t'} \in SEM^{reify}$ is a reified semantic relation, $t'$ is of the form $\sigma^i(t_1, \ldots, t_i)$, and there is a rule of the form $Sem_A(t', \overline{x}') \leftarrow \varphi \in rules$, then $Sem_A^{t'}(\overline{x}') \leftarrow \varphi[Sem_{A_j}(t_j, \overline{x}_j) \mapsto Sem_{A_j}^{t_j}(\overline{x}_j)] \in rules^{reify}$ is a reified semantic rule, and

(3) if $Sem_A^{t'}(\overline{x}') \leftarrow \varphi \in rules^{reify}$ is a reified semantic rule and $Sem_{A_j}^{t_j}$ appears in $\varphi$, then $Sem_{A_j}^{t_j} \in SEM^{reify}$ is a reified semantic relation.

THEOREM 5.3 (REIFICATION IS SOUND). *Given a formula $\varphi$ and a set, rules, of semantic rules, let $\psi$ be the formula in which every occurrence of $Sem_A(t, \overline{x})$ is replaced by the reified semantic relation $Sem_A^t(\overline{x})$, and $rules^{reify}$ is the conjunction of the reified semantic rules produced by REIFY(rules, $Sem_A$, $t$) for each $Sem_A(t, \overline{x})$ appearing in $\varphi$. The constraint $\varphi$ is valid under the original semantic rules rules if and only if $\psi$ is valid under the reified semantic rules: $rules \models \varphi \Leftrightarrow rules^{reify} \models \psi$.*

*Semantic Relation Inlining.* In general, the MuVal solver that we use to solve $\mu$CLP queries scales poorly in the number of relations used to define the semantics of a program. The goal of the optimization INLINE is to eliminate semantic relations by inlining their definitions.

The SMT encoding SMT-FORMULA-OF in §2.1 can be seen as an application of INLINE that eliminates semantic relations by inlining their meaning into a quantified first-order formula (i.e. $\varphi_{Sem_{max2}}$ in §2.1). The optimization INLINE applies a similar insight to as many semantic relations as possible to reduce the number of semantic relations that the final solver has to deal with.

The inlining optimization INLINE is implemented nearly identically to the SMT-FORMULA-OF encoding defined in §4.1, except the condition on line 7 of SMT-FORMULA-OF is changed to ensure termination by stopping iteration if it comes to a semantic rule that is self-recursive on $t'$.

*Example 5.4.* Continuing Example 5.1, semantic-inlining inlines $Sem_S^{x--}$ and $Sem_S^{x--;y++}$ to yield $Sem_S^{x--;\ y++}(x, y, x', y') \leftarrow \exists x'', y''. x'' = x - 1 \wedge y'' = y \wedge x' = x'' \wedge y' = y'' + 1$.

*Quantifier Elimination.* In the above example, we see that inlining definitions can leave superfluous quantifiers (e.g. $\exists x'', y''$. in $Sem_S^{x--;\ y++}$). Eliminating these unnecessary quantifiers using simple quantifier-elimination methods can yield formulas that are easier for existing solvers to handle, which often exhibit performance that degrades exponentially in the number of quantifier alternations. Continuing the above example, in MUSE, we apply Z3's quantifier-elimination tactic rule-by-rule to yield the quantifier-free formula $Sem_S^{x--;\ y++}(x, y, x', y') \leftarrow x' = x - 1 \wedge y' = y + 1$.

## 6 EXPERIMENTS

We evaluated MUSE with respect to the following research questions:

**Q1:** How effective is MUSE at verifying solutions to SemGuS$^\mu$ problems?
**Q2:** How effective are the optimizations from §5 at improving MUSE's performance?
**Q3:** Does MUSE enable SemGuS synthesizers to handle problems with logical specifications?

All experiments were conducted on a desktop running Ubuntu 18.04 LTS, equipped with a 4-core Intel(R) Xeon(R) processor running at 3.2GHz with 12GB of memory. For each experiment we allotted a maximum of 6GB of memory; for each verification query we set a timeout of 5 minutes; and for each synthesis query we set a timeout of 30 minutes. We repeated each experiment three times and report the median result.

### 6.1 Benchmarks

We collected 141 SemGuS problems—whose semantics and logical specifications were expressed within linear integer arithmetic—from the official SemGuS benchmarks (https://github.com/SemGuS-git/Semgus-Benchmarks)—which we augmented with quantified logical specifications.

Verifying Solutions to Semantics-Guided Synthesis Problems 17

Table 1. All benchmarks, broken down by benchmark suite. For each solver, we list the number of benchmarks to which it can be applied, the number of benchmarks solved, and the average time per solved instance. The results given in the table are for the best configuration of each solver. The rows for "Partial Correctness" and "Total Correctness" summarize the results for all SemGuS benchmarks that require proving/refuting partial and total correctness, respectively, of imperative programs containing a loop (across all benchmark suites).

| Suite | Total | SMT-FORMULA-OF | | | CHC-OF | | | MUCLP-OF | | Best | |
|---|---|---|---|---|---|---|---|---|---|---|---|
| | | isSMT | # Verified | Time | isCHC | # Verified | Time | # Verified | Time | # Verified | Time |
| SyGuS | 160 | 160 | 104 | 0.42s | 160 | 152 | 0.18s | 111 | 15.89s | 152 | 0.12s |
| SyGuS-Imp | 54 | 27 | 27 | 0.07s | 54 | 54 | 2.27s | 28 | 1.32s | 54 | 2.04s |
| FuncImp | 38 | 12 | 12 | 0.07s | 11 | 11 | 0.13s | 35 | 4.26s | 35 | 1.02s |
| Boolean | 176 | 176 | 176 | 0.08s | 176 | 176 | 0.09s | 169 | 2.44s | 176 | 0.08s |
| Regex | 102 | 92 | 77 | 4.09s | 102 | 96 | 4.61s | 46 | 4.83s | 97 | 3.28s |
| ScoreCards | 60 | 60 | 40 | 0.10s | 60 | 60 | 0.43s | 51 | 3.69s | 60 | 0.10s |
| Partial Correctness | 70 | 27 | 27 | 0.07s | 70 | 63 | 1.96s | 40 | 2.20s | 70 | 1.76s |
| Total Correctness | 18 | 0 | – | – | 0 | – | – | 15 | 5.34s | 15 | 5.34s |
| Controllers | 80 | 0 | – | – | 0 | – | – | 75 | 5.50s | 75 | 5.50s |
| PDDL | 20 | 0 | – | – | 0 | – | – | 20 | 2.37s | 20 | 2.37s |
| Games | 40 | 0 | – | – | 0 | – | – | 40 | 0.66s | 40 | 0.66s |
| Total | 730 | 527 | 436 | 0.86s | 563 | 549 | 1.15s | 575 | 5.67s | 709 | 1.40s |

Because of the limitations of MuVal [40] (the solver we use for reasoning about $\mu$CLP queries), we restricted our focus to examples whose semantics can be expressed in LIA. We additionally created 80 SemGuS problems encoding SyGuS problems from SyGuS-comp (https://github.com/SyGuS-Org/benchmarks). We created an additional 74 SemGuS problems and 70 SemGuS$^\mu$ problems. Our suite of benchmarks consists of 295 functional-synthesis problems and 70 reactive-synthesis problems. It consists of SemGuS problems for various domains, including imperative programs, functional programs, score cards, SyGuS problems, Boolean formulas, regular expressions. Our SemGuS$^\mu$ problems consider relatively small (but infinite-state) reactive-synthesis problems, including reactive controllers, LTL formulas, reachability and Büchi games, and robotic path-planning problems. We split our problems into 9 suites (cf. Table 1). A complete description of the benchmarks can be found in Appendix B, including a description of each suite of benchmarks, the sizes of benchmarks, and the process by which we generated each specification.

### 6.2 Experimental Setup

We evaluate Muse as follows: (*i*) we ablate each optimization described in §5 to evaluate Muse's performance on hand-crafted verification problems, and (*ii*) we incorporate Muse into an existing SemGuS synthesizer, Ks2 [26], to determine if Muse enables solving SemGuS problems with logical specifications. For both studies, we evaluate on the suite of SemGuS problems described in §6.1.

*Ablation Study.* For each SemGuS problem described in §6.1, we handcrafted two programs: one that satisfies the SemGuS problem, and one that does not. To generate the hand-crafted programs, we created the smallest program that satisfied the SemGuS problem. We then made minimal modifications to the satisfying program so that it no longer satisfied the logical specification. For each benchmark, we ran the SMT-FORMULA-OF, CHC-OF, and MUCLP-OF based solvers. We ran each solver in five configurations: (*i*) without any optimizations, (*ii*) with all three optimizations, (*iii*) with reification and predicate inlining, (*iv*) with reification and quantifier elimination, and (*v*) with predicate inlining and quantifier elimination. Figure 4 compares each configuration for each solver, and Table 1 details the result of the best configuration for each solver.

*SemGuS Synthesizer Study.* Prior to Muse, most SemGuS synthesizers have limited their scope to solving problems whose specifications are represented as a finite set of input-output examples. The SemGuS toolkit [26] introduced Ks2, which was the first SemGuS synthesizer that was in theory capable of handling SemGuS problems with logical specifications. However, the baseline verifier provided with Ks2 relies on a naive SMT encoding that uses recursive axioms, a theory that is not well supported by SMT solvers. As such, one of our goals in developing a general-purpose verifier for SemGuS is to enable Ks2 (and other SemGuS synthesizers) to solve SemGuS problems with logical specifications. We incorporated MUSE into Ks2 to replace the baseline solver to determine if MUSE would allow Ks2 to solve more SemGuS problems with logical specifications.

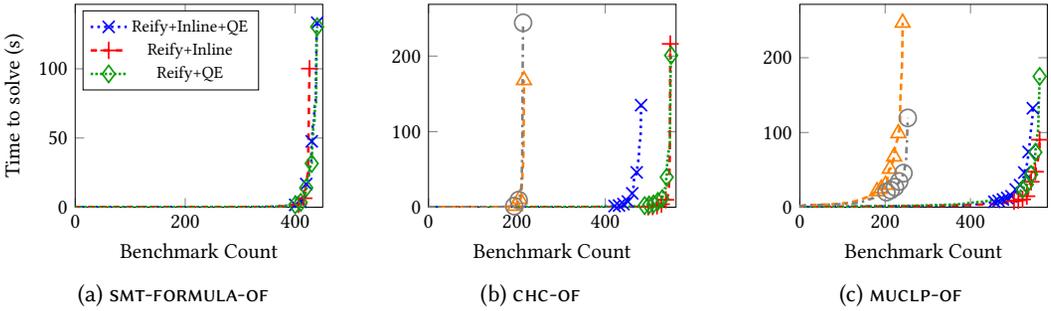

Fig. 4. Three cactus plots, one per encoding, with one line per optimization configuration used. If a point $(x, y)$ appears along the line labeled by *config*, then *config* solved $x$ instances in under $y$ seconds. A line lower and further to the right is better.

Ks2 is a top-down enumeration-based SemGuS synthesizer that is available as part of the SemGuS toolkit (a collection of tools made available to enable development of SemGuS synthesizers) as a baseline synthesizer [26].[2] At a high-level, Ks2 operates by generating terms, checking each term against a set of input-output examples, then verifying if the term is correct with respect to the logical specification (if provided) and returns the first term that satisfies the specification. We modified Ks2 to use our verifier MUSE instead of the baseline verifier when checking if enumerated terms meet a logical specification. We denote the augmented synthesizer as Ks2+MUSE. For each verification problem encountered, we had Ks2+MUSE invoke the simplest solver (i.e., SMT-FORMULA-OF for non-recursive solutions, CHC-OF for CHC-like problems, and MUCLP-OF for all others) with reification and predicate inlining (the configuration that performed the best in the ablation study).

We evaluated Ks2+MUSE against baseline Ks2 on the 295 SemGuS problem described in §6.1. We excluded the reactive-synthesis problems because Ks2 does not support SemGuS$^\mu$ problems (i.e., problems whose semantics are not represented as CHCs). For each SemGuS problem, we provided 5 input-output examples alongside the logical specification to improve the search performed by Ks2.

Table 1 details the results of the MUSE ablation study. Table 2 details the results of the synthesis study. While both Table 1 and Table 2 are based on the same set of benchmarks, *Table 1 and Table 2 consider a different number of experiments*—i.e., Table 1 considers 730 verification experiments (one satisfying and one falsifying program for each of the 295 SemGuS and 70 SemGuS$^\mu$ benchmarks), while Table 1 summarizes the results of running Ks2+MUSE on each of the 295 SemGuS benchmarks.

### 6.3 Q1: Effectiveness of MUSE

Table 1 presents the results of our experiments, summarized per benchmark suite, and summarizes how the three encodings supported by MUSE compare. Each column labeled by a variant of MUSE indicates the number of instances solved within the allotted time limit, together with the average solving time. The first columns of the SMT-FORMULA-OF and CHC-OF blocks also specify how many instances fall within the fragment of first-order logic to which they can be applied. We note that the MUCLP-OF variant solved the most instances, solving 575 instances taking on average 5.6 seconds.

---
[2]Ks2 is available as part of the SemGuS toolkit from the official SemGuS website: https://www.semgus.org.

The CHC-OF variant came in a close second solving 549 instances, averaging 1.1 seconds each. Upon further investigation, we found that CHC-OF solved 55 instances not solved by the other two variants, while MUCLP-OF solved 145 instances that were not and *could* not be solved by the other variants. Overall, 709 of the 730 benchmarks were solved by at least one of the three solvers. For the benchmarks that all solvers could handle, CHC-OF and SMT-FORMULA-OF performed similarly and were 12X faster than MUCLP-OF on average (geomean). For each problem for which a solver terminated for both the questions of verifying a correct solution and refuting an incorrect solution, proving that a solution was correct was in general slower than proving that a solution was incorrect (avg. 1.2X slower for CHC-OF, 2.8X slower for MUCLP-OF, 1.1X slower for SMT-FORMULA-OF).

MUSE solved 709/730 verification instances (in at least one variant). We inspected the instances where MUSE failed: across our 295 SemGuS problems (590 verification queries), 88 verification queries required reasoning about a program containing a loop, 70 of which required proving/refuting partial correctness and 18 of which required proving/refuting total correctness. Our benchmarks skew more towards partial correctness because the SyGuS-IMP benchmark includes 54 such verification problems that all involve partial-correctness specifications. While MUSE could solve all of the partial-correctness queries, it had a harder time proving/refuting total correctness, failing to refute one program and failing to prove totally correct 2/9 programs. Of the remaining 13 unsolved SemGuS verification queries (from the Regex and SemGuS-encoding-of-SyGuS problems), the verification query required proving correctness for programs whose semantics were relatively large (over 1000 AST nodes, some of which had over 500,000 AST nodes).

For SemGuS$^\mu$ verification queries, MUSE solved 135/140 of them. We note that due to the added complexity of encoding reactive synthesis problems within SemGuS, many of our reactive synthesis benchmarks consider relatively small (but infinite state) reactive synthesis problems, for which we expect existing techniques specialized for reactive synthesizers would excel.

To answer **Q1**, the verification techniques implemented in MUSE are effective and can verify 709/730 of our problem instances. We found that overall, proving correctness of a solution is generally harder than proving that a solution is incorrect. We note that in its current state MUSE is limited in how it can be incorporated into synthesis algorithms (e.g., those based on CEGIS [1]) because there is not yet a principled way to extract a meaningful counter-example when a solution fails to satisfy a SemGuS$^\mu$ problem. The root cause of this limitation is that MuVal does not provide a counter-model/proof when it refutes a formula.

### 6.4 Q2: Effectiveness of the Optimizations from Section 5

Figure 4 illustrates how the optimizations described in §5 affect the performance of each encoding. Each of the three graphs in Figure 4 shows the results of an ablation study—i.e., we compared the effectiveness of the optimizations by considering five configurations: no optimizations, all optimizations, and three configurations in which a single optimization was disabled. The three cactus plots show the performance for the three encodings in §4; the lines in each cactus plot show the performance of the five considered optimization configurations. One solver is better than another if its line is lower and to the right of the other solver's line (i.e., it can solve more problems in less time). For example, the SMT-FORMULA-OF variant performed best using reification and quantifier elimination. For all three encodings, the configurations using reification performed the best, solving on average 2.5X the number of verification problems solved without reification. In-lining and quantifier elimination do help somewhat; however, closer inspection of the results revealed that attempting quantifier elimination on the larger formulas generated during the execution of SMT-FORMULA-OF can lead to poor results.

To answer **Q2**, the optimizations are very effective, with reification being the most effective.

Table 2. Results of running Ks2+Muse on SemGuS benchmarks.

| | Suite | SyGuS | SyGuS-Imp | FuncImp | Boolean | Regex | ScoreCards | Total |
|---|---|---|---|---|---|---|---|---|
| Ks2+Muse | Solved | **10** | **25** | **10** | **38** | **33** | **19** | **135** |
| | Timed Out | 1 | 2 | 10 | 5 | 10 | 1 | 29 |
| | Memed Out | 54 | 0 | 0 | 45 | 8 | 10 | 117 |
| | Verification Calls | 1210 | 367 | 53 | 38 | 108 | 19 | 1795 |
| | Time (s) | 476.46 | 3319.77 | 48.14 | 984.53 | 1240.85 | 58.19 | 6118.94 |
| | Verification Time (s) | 453.33 | 3074.16 | 24.02 | 16.60 | 197.16 | 8.62 | 3767.89 |
| | Verif. Time / Call (s) | **0.37** | **8.38** | **0.45** | **0.43** | **1.82** | **0.45** | **2.10** |
| Ks2 | Solved | 3 | 0 | 4 | 35 | 0 | **19** | 61 |
| | Timed Out | 62 | 27 | 16 | 53 | 51 | 11 | 234 |
| | Memed Out | 0 | 0 | 0 | 0 | 0 | 0 | 0 |
| | Verification Calls | 81 | – | 19 | 35 | – | 19 | 154 |
| | Time (s) | 30.74 | – | 105.49 | 9749.43 | – | 250.29 | 10135.95 |
| | Verification Time (s) | 29.81 | – | 81.50 | 9554.98 | – | 190.17 | 9856.46 |
| | Verif. Time / Call (s) | **0.37** | – | **4.24** | **273.00** | – | **10.00** | **64.00** |

## 6.5 Q3: Integration with an Enumeration-Based SemGuS Synthesizer

Table 2 summarizes the results of the synthesis study. We found that by incorporating Muse in Ks2, we enabled Ks2+Muse to solve 135 SemGuS problems with logical specifications—*74 of which could not be solved by Ks2 with the baseline verifier, nor any other previous SemGuS synthesizer*. For SemGuS problems that Ks2+Muse did solve, Ks2+Muse required on average 13.3 verification calls, taking on average 2.1 seconds each. Instead, Ks2 with the baseline verifier solved only 61 SemGuS problems, taking on average 2.5 verification calls, each taking on average 64 seconds. We found that neither Ks2+Muse nor baseline Ks2 solved many of the SemGuS-encoding-of-SyGuS problems; however, upon closer inspection we found that Ks2+Muse typically ran out of memory while enumerating terms whereas the baseline Ks2 timed out during a verification call. We note that all 65 of the SemGuS-encoding-of-SyGuS problems (when expressed as a SyGuS problem) could be solved by the CVC5 SyGuS solver [5] in under a second. As such, if one knows that a synthesis problem can be expressed in SyGuS, then using a SyGuS solver is obviously better.

For both synthesizers, we inspected the SemGuS problems that could not be solved within the time limit. Among the 146 SemGuS problems that Ks2+Muse could not solve, the 55 SemGuS-encoding-of-SyGuS problems failed to enumerate the correct solution within the time limit. Similarly, 1 FuncImp, 50 Boolean, 10 Regex, and 11 Scorecard problems failed to enumerate a correct solution within the time limit. The remaining 8 FuncImp, 2 SyGuS-Imp, and 8 Regex benchmarks timed out during verification after enumerating the correct solution. For the majority of the 234 SemGuS problems not solved by the baseline Ks2, Ks2 spent the majority of its time in the baseline verifier.

To answer **Q3**, by incorporating Muse into Ks2, we enabled Ks2 to synthesize 135 verifiably correct solutions, 74 of which could not be solved by any other SemGuS synthesizer. Of the 146 problems that remained unsolved, only 18 timed out due to verification (of which 11 could be solved by the virtual best solver, which runs each solver in parallel and returns once any terminates).

## 7 RELATED WORK

*Fixed-Semantics Program Verification.* There is a large volume of work on automated verification for programs within a fixed language semantics. The typical approach often depends on the form of the language considered. For example, a popular verification methodology for imperative programs is to generate verification conditions automatically, and use invariant-generation techniques to satisfy those conditions [21, 24, 31, 34]. For functional languages, the typical approach uses type-based reasoning [35, 39, 42]. While there are a plethora of techniques for verification of imperative, declarative, and functional languages, these approaches (unlike Muse) do not support a user-defined semantics (e.g., via SemGuS$^\mu$), but can use domain-specific techniques to obtain better performance.

*Verification Frameworks.* More closely related to our work is the line of work on verification frameworks and intermediate verification languages. Stefănescu et al. [38] describe how to create an automated program verifier for arbitrary languages automatically, from an *operational* semantics written in the K framework. In a conversation with the K team, we confirmed that while matching logic—an expressive first-order fixed-point logic similar to $\mu$CLP—forms the basis of the K framework, the produced verifier is limited to answering reachability queries on inductively defined languages [32]. For example, while one can write a matching-logic formula that encodes the Büchi game in Figure 3, the K framework does not support defining a language whose semantics is a Büchi game (i.e., one cannot automatically generate a verifier for the language). Similarly, the verifier produced by K for an encoding of the IMP language in Figure 2 would not be able to verify that $s_{loop}$ is totally correct (cf. §2.2). Finally, while the verifier automatically produced by the K framework cannot answer these verification problems, matching logic is general enough to express formulas with both least and greatest fixed-points. A reduction to matching logic might serve as the basis for yet another verifier for SemGuS.

In a similar vein is work on creating and using an intermediate language for verification, such as Boogie [7] or Why3 [17]. A key difference between those tools and Muse is that with Boogie and Why3, a language's semantics is specified via a *translational semantics*, whereas with Muse a language's semantics is specified *declaratively*. With a translational semantics, one has to define a function that walks over the abstract-syntax tree $t$ of a program, and constructs an appropriate Boogie/Why3 program whose meaning captures the semantics of $t$. In contrast, with Muse, the semantics is specified declaratively, using logical relations in SemGuS$^\mu$, thus allowing one to model many diverse scenarios (e.g., our robot example). Many systems have used Boogie as their intermediate language, including Dafny [31] and VCC [14]. Other similar systems include Cameleer [35] (on top of Why3) and various C analyzers built on top of the FRAMA-C platform [29]. While intermediate verification languages allow the reuse of verifiers for multiple languages, they generally support a single *language paradigm* (e.g, object-oriented, functional, etc.) and a single verification strategy (e.g., pre/post conditions and loop invariants) and thus may be difficult to use for a language based on a different paradigm. In contrast, the SemGuS$^\mu$ framework uses a logic-based approach to specifying semantics, which allows Muse to be applied to a wide variety of problems.

*Logic-Based Verification.* There is also a substantial body of work that uses fragments of first-order logic (with fixed-points) to verify programs. A broad class of work considers programs represented as transition formulas [2, 4, 6, 16]. That is, the verification task takes as input a formula modeling the transitions of the program. While these techniques and ours all take as input a logical formalism describing the program of interest, transition formula are monolithic formulas defined on a program-by-program basis, and differ from the modular semantic relations—supplied on a per-language basis via SemGuS$^\mu$—used in Muse.

Our work on Muse also has connections to verification based on answer-set programming and stratified logic programming [12, 13, 18]. Specifically, techniques based on answer-set programming and stratified logic programming automatically compile a program and specification into a logic program. Our technique could be viewed similarly, specifically because $\mu$CLP can be seen as a generalization of logic programs (with first-order constraints and fixed-point operators). While the SemGuS$^\mu$ framework expects a user to provide a declarative specification of a language's semantics, typical approaches based on answer-set programming or stratified logic programming expect the user to provide an encoding of both the specification and the program's semantics on a program-by-program basis.

Another line of work translates verification queries into validity (or satisfiability) queries in fragments of first-order logic [23, 41]. SeaHorn [23] compiles annotated C programs into a system of CHCs to discharge the generated verification conditions. The work of Unno et al. [41] formulates a number of program-verification tasks in the pfwCSP fragment of first-order logic (a constraint language similar in expressiveness to $\mu$CLP). These techniques are similar to the ones used in Muse in that they answer verification queries by generating a logical query and using a solver to answer the query; however, the methods in these other tools are defined for a fixed language (e.g., a fragment of C), whereas Muse is parameterized by the language specified in the SemGuS$^\mu$ input.

*SemGuS Synthesizers.* As discussed in §6, we incorporated our SemGuS verifier, Muse, within an existing SemGuS synthesizer Ks2, which enabled Ks2+Muse to solve 135 SemGuS problems with logical specifications—74 of which could not be solved by any previous SemGuS synthesizer. In addition to describing Ks2, Johnson et al. [26] also present a format for describing SemGuS problems, which our tool Muse accepts. Besides Ks2, the only other existing SemGuS synthesizer is Messy, which only handles specifications in the form of input-outuput examples [28]. While Messy can prove if a SemGuS problem is realizable/unrealizable, when Messy proves that a SemGuS problem is realizable, it is unable to produce a concrete program.[3] For this reason, we could not combine our verifier Muse with Messy to enable it to solve SemGuS problems with logical specifications.

While Ks2 is the first SemGuS synthesizer capable of solving SemGuS problems with logical specifications, there are other domain-specific synthesizers that can (quickly) solve some of the synthesis problems used in our benchmarks when represented within the constraints of their domains. For example, CVC5 [5] can quickly solve all of the SyGuS problems that we encoded within the SemGuS framework to evaluate Ks2+Muse. However, to solve a SemGuS problem using a SyGuS solver, we would first need to recognize that the SemGuS problem encodes a SyGuS problem—i.e., by proving that the syntax and semantics of the SemGuS problem can be reformulated automatically as a SyGuS problem. To summarize, to support a general framework, we must develop general SemGuS solvers.

*Reactive Synthesis.* Our work on SemGuS$^\mu$ has ties to reactive synthesis [3, 11, 37]—i.e., SemGuS$^\mu$ is flexible enough to encode both functional and reactive-synthesis problems (e.g., Büchi and reachability games, controllers, and imperative programs that satisfy a temporal property). However, although SemGuS$^\mu$ is general enough to encode such problems, and thus offers a way to reason uniformly about functional and reactive-synthesis problems, we do not expect solvers based on Muse to be competitive compared to specialized reactive synthesizers (e.g., [11, 33, 37]).

## 8 CONCLUSION

The SemGuS framework [28] is becoming a standard for program synthesis, as happened with the less expressive SyGuS framework. This paper presents the methodology for verifying that a candidate solution is valid for a SemGuS problem. Our technique reduces verification questions to validity questions in $\mu$CLP (a fragment of first-order fixed-point logic), or validity questions in easier logics, when possible. Our work fills an important gap in the pipeline of techniques needed to build practical SemGuS synthesizers. One can now build solvers that synthesize solutions to SemGuS problems involving complex specifications and verify whether these solutions are correct! While our tools currently handle relatively small programs, improvement to our framework will lead to improvements in any SemGuS solver.

## ACKNOWLEDGMENTS

Supported, in part, by a Microsoft Faculty Fellowship; a UCSD JSOE Scholarship; a gift from Rajiv and Ritu Batra; and NSF under grants CCF-1750965, CCF-1918211, CCF-2023222, CCF-2211968, and CCF-2212558. Any opinions, findings, and conclusions or recommendations expressed in this publication are those of the authors, and do not necessarily reflect the views of the sponsoring entities.

---

[3]This limitation is detailed in the official documentation of Messy: https://github.com/SemGuS-git/Semgus-Messy.

## A EXTENDED OVERVIEW

In §2, we detail a variety of example SemGuS verification problems to explain our technique of reducing SemGuS verification queries into validity queries in various fragments of first-order logic. Specifically, this section extends §2 by detailing another interesting class of SemGuS verification problems—problems whose verification queries can be encoded into a decision problem over co-CHCs (the DeMorgan duals of CHCs). We further demonstrate a methodology for splitting specifications—i.e., splitting a specification into a logically equivalent conjunction of sub-specifications and independently solving the SemGuS verification problem induced by each sub-specification.

### A.1 DoubleViaLoop Total: Quasi Co-CHC and Specification Splitting

In §2.2, we were able to use a CHC solver to reason about the query $Q_{loop}$ because the specification $\varphi_{loop}$ did not contain positive occurrences of the semantic relations in $Sem_{loop}$ and when the specification did contain positive occurrences of the semantic relations in $Sem_{loop}$ we could encode the verification query into a verification query in the $\mu$CLP calculus. In the next example, we consider the same grammar $G_{loop}$ and semantics $Sem_{loop}$ as in Figure 2, but introduce a modified specification $\varphi_{loop}^{tot}$ that requires a form of total correctness:

$$\varphi_{loop}^{tot} = \forall x, y'.(0 \leq x \land 2x = y') \Rightarrow Sem_L(f_{loop}, x, 0, 0, y') \tag{5}$$

The above specification states that when the program to be synthesised starts in a state where variable x takes a non-negative value $x$ and variable y is 0, then it will terminate in a state where x is 0 and y is twice x's initial value $x$. However, the resulting query $Q_{loop}^{tot} = Sem_{loop}^{LFP} \models \varphi_{loop}^{tot}[f_{loop} \mapsto s_{loop}]$ cannot be solved via a reduction to CHC satisfiability because the specification has a positive occurrence of the semantic relation $Sem_L$. Instead, we show that one *can* construct a logically equivalent query $\overline{Q_{loop}^{tot}}$ that can be solved via a reduction to co-CHC falsafiability. Whereas SemGuS semantics are defined as the least solution of a set of CHCs, the semantics could instead be equivalently defined by the greatest solution to a set of co-CHCs (that are defined dually to the set of CHCs). This is achieved by defining a new relation—the *complement relation* $\overline{Sem}_L(floop, x, 0, 0, y')$—as a co-CHC, which allows us to explicitly reason about negative occurrences of $Sem_L$. This approach allows us to solve a new query in which all relations (*i*) are defined as greatest fixed-points, and (*ii*) appear negatively within the specification:

$$\overline{Q_{loop}^{tot}} \triangleq \overline{Sem}_{loop}^{GFP} \models \forall x, y'.(0 \leq x \land 2x = y') \Rightarrow \neg \overline{Sem}_L(s_{loop}, x, 0, 0, y'), \tag{6}$$

where $\overline{Sem}_{loop}^{GFP}$ is the greatest solution to the co-CHCs that define the dual semantics of $Sem_{loop}$. The semantics $Sem_{loop}$ explicitly describes the behaviors each program can exhibit—e.g., $Sem_L(t, x, y, x', y')$ states that, on input state $\langle x, y \rangle$, the program $t$ *can* terminate with the output state $\langle x', y' \rangle$. The dual semantics instead describes the complement of $Sem_{loop}$—i.e., the behaviors that each program cannot exhibit. For example, $\overline{Sem}_L(t, x, y, x', y')$ states that on input $\langle x, y \rangle$ the program $t$ *cannot* terminate with the output state $\langle x', y' \rangle$—i.e., for every execution of $t$ on the input $\langle x, y \rangle$, $t$ either does not terminate or $t$ terminates in a state different than $\langle x', y' \rangle$. Thus, the query $\overline{Q_{loop}^{tot}}$ asks if there is a positive value for $x$ for which the candidate program does not compute $2x$—either because the candidate program does not terminate on the input or because it terminates in a state where y is not $2x$.

Muse takes as input $s_{loop}$, $Sem_{loop}$, and $\varphi_{loop}^{tot}$ and produces the query $\overline{Q_{loop}^{tot}}$. The process dualizes every semantic relation and each semantic rule. For example, our encoding produces the following rule to define $\overline{Sem_S}(\text{s; t}, x, y, x', y')$.

$$\overline{Sem_S}(\text{s; t}, x, y, x', y') \Rightarrow \forall x'', y''. \overline{Sem_S}(\text{s}, x, y, x'', y'') \vee \overline{Sem_S}(\text{t}, x'', y'', x', y')$$

After the encoding, the produced query $\overline{Q_{loop}^{tot}}$ is logically equivalent to the original query $Q_{loop}^{tot}$. We note that the standard decision problem for coCHCs is satisfiability—which is equi-expressive to CHC satisfiability. In fact, the standard approach to coCHC satisfiability is to dualize the set of coCHCs (to produce a set of CHCs) check satisfiability of CHCs and complement the final result. As such, the query $\overline{Q_{loop}^{tot}}$ cannot be solved via a reduction to coCHC satisfiability. Instead, we assume the decision problem of interest is coCHCs falsification (the dual problem of satisfiability)—is the given query falsified by every interpretation of the set of co-CHCs. However, we are not aware of any solvers specialized in solving CHC falsification. Thus in Muse we encode $\overline{Q_{loop}^{tot}}$ as an equivalent µCLP problem and solve using the µCLP solver MuVal [40] to solve this verification query in 1.6 seconds.

*Specification Splitting.* In §2.2 and in the above example, we considered solutions that could reduce the verification query to either CHC satisfiability or co-CHC falsification because the semantic relations appear in the specification either *only* positively (in which case we use the dual semantics) or *only* negatively (in which case we use the original semantics).

Expressing total correctness requires specifications in which the semantic relations appear both positively and negatively, as in the following example:

$$\varphi_{loop}^{both} = \forall x, y'. 0 \le x \Rightarrow (Sem_L(x, 0, 0, y') \Leftrightarrow 2x = y') \tag{7}$$

Although this specification contains both positive and negative occurences of semantic relations and cannot be reduced to either CHC satisfiability or co-CHC falsification, the specification can be split into two separate specifications in which the semantic relations appear only positively in one, and only negatively in the other. In fact, $\varphi_{loop}^{both}$ is logically equivalent to the conjunction of the specifications $\varphi_{loop}$ and $\varphi_{loop}^{tot}$. To verify that a candidate program satisfies the specification in $\varphi_{loop}^{both}$, it is sufficient to check that the candidate program satisfies both $\varphi_{loop}$ and $\varphi_{loop}^{tot}$ by checking validity of the queries $Q_{loop}$ and $\overline{Q_{loop}^{tot}}$.

## A.2 Encoding CHC-like SemGuS$^\mu$ Verification problems with coCHCs

In §4.2, we defined CHC-OF that automatically reduced SemGuS$^\mu$ verification problems whose semantics were CHC-like and whose specification contained only negative occurrences of semantic relations into a CHC satisfiability problem. In this section, we define the procedure CO-CHC-OF which is similarly defined to encode SemGuS$^\mu$ verification problems whose semantics are CHC-like and whose specification contains only positive occurrences of the semantic relations into a co-CHC falsification problem. A co-CHC is a formula of the form:

$$\forall \bar{x}_0, \ldots, \bar{x}_n. R_0(\bar{x}) \Rightarrow R_1(\bar{x}_1) \vee \cdots \vee R_n(\bar{x}_n) \vee F(\bar{x}_0, \ldots, \bar{x}_n),$$

where each component is as described when defining CHCs (cf. Equation (4)). Note that the definitions presented here are logically equivalent to the typical definition used in constraint logic programming [40].

The procedure CO-CHC-OF is nearly identical to CHC-OF. The procedure uses the auxiliary function $dual(\varphi) = \neg \varphi[Sem_A(t, \Gamma, \Upsilon) \mapsto \neg \overline{Sem_A}(t, \Gamma, \Upsilon)]$ that computes the dual of the input formula (i.e., if $\varphi$ is a CHC then $dual(\varphi)$ is a co-CHC). The CO-CHC-OF procedure changes lines 2 and 9. Line 3 becomes $\psi \leftarrow dual(\varphi[\langle f, A \rangle \in F.f \mapsto P(f)])$ and line 9 becomes $rules \leftarrow rules \wedge \bigwedge \{dual(rule) : rule \in rules'\}$.

Finally, if the specification $\varphi$ contains both positive and negative occurrences of semantic relations, but can be split into two specification $\varphi^+$ and $\varphi^-$ that, respectively, contain only positive and only negative occurrences of semantic relations, then $\varphi$ can be encoded into two separate problems using the CHC-OF and CO-CHC-OF encodings.

## A.3 Hyperproperties: $\mu$CLP

§2.2 presented a technique for verifying a solution for cases when the specification contains only negative occurrences of the semantic relations, and Appendix A.1 showed a solution for cases when the specification contains only positive occurrences of the semantic relations and even how to split a specification into sub-specifications with only positive or only negative occurrences of the semantic relations. However, not all specifications can be split this way.

For example, consider the following specification that requires the synthesized function to be commutative in its arguments. (Such properties are sometimes called *hyperproperties* because their falsification requires one to consider two different executions of the program, starting from different input states.)

$$\varphi_{comm} \triangleq \forall x, y, x', y'. Sem_L(s, x, y, x', y') \Rightarrow Sem_L(s, y, x, x', y') \tag{8}$$

The specification $\varphi_{comm}$ could arise when trying to synthesize a program like

$$s_{plus} = \texttt{while 0 < x do x--; y++} \tag{9}$$

which, when it terminates, sets the value of variable y to the sum of the inputs x and y. This program is in the language defined by the grammar $G_{loop}$ in Figure 2a, and we assume it operates over the semantics $Sem_{loop}$ in Figure 2d.

To prove that the program $s_{plus}$ satisfies the specification $\varphi_{comm}$, one must simultaneously reason about the the positive and negative occurrences of $Sem_L$ in $\varphi_{comm}$. We show that the problem of verifying whether a program meets a specification like $\varphi_{comm}$—and in fact every specification expressible in SemGuS—can be reduced to checking validity in the $\mu$CLP calculus, a logic that combines least- and greatest-fixed-point reasoning [40] (cf. §4.3 for details).

MUSE reduces this verification problem to a $\mu$CLP query equivalent to the following verification query that combines the semantics $Sem_{loop}$ with its dual $\overline{Sem}_{loop}$ (the dual is computed as the conjunction of the DeMorgan dual of each rule in $Sem_{loop}$):

$$Q_{comm} \triangleq Sem_{loop}^{LFP} \wedge \overline{Sem}_{loop}^{GFP} \models \forall x, y, x', y'. \overline{Sem}_L(s_{plus}, x, y, x', y') \vee Sem_L(s_{plus}, y, x, x', y'), \tag{10}$$

which follows from Equation (8) by (*i*) instantiating $s$ as $s_{plus}$, (*ii*) replacing "$Sem_L(s_{plus}, x, y, x', y') \Rightarrow$ ..." with "$\neg Sem_L(s_{plus}, x, y, x', y') \vee$ ...," and (*iii*) replacing "$\neg Sem_L(s_{plus}, x, y, x', y')$" with "$\overline{Sem}_L(s_{plus}, x, y, x', y')$." Recall that $Sem_{loop}^{LFP}$ is defined as the least solution to the CHCs defining $Sem_{loop}$. We similarly define $\overline{Sem}_{loop}^{GFP}$ as the greatest solution to the co-CHCs defining $\overline{Sem}_{loop}$.

In our tool MUSE, we use the $\mu$CLP solver MuVal [40] to solve $Q_{comm}$ in 6.2 seconds, thereby proving that $s_{plus}$ is commutative. In general, $\mu$CLP allows for both positive and negative occurrences of predicate variables (i.e., semantic relations) within the query (i.e., specification), which makes the use of both the original and dual semantics redundant. However, MuVal—the underlying $\mu$CLP solver we use within MUSE—only allows for queries with positive occurrences of predicate variables. Thus, we appropriately transform our verification query to transform negatively occurring semantic relations into positively occurring predicate variables by making use of the dual semantics when encoding the verification problem into $\mu$CLP.

## B  BENCHMARK DESCRIPTIONS

This section provides a detailed description of each suite of benchmarks listed in Table 1. Furthermore, we provide details about the average size and complexity of each SemGuS verification and synthesis problem we consider. Finally, we explain the process by which we generated logical specifications for each of the existing 141 SemGuS problems.

*SyGuS.* The SyGuS suite consists of 80 SemGuS encodings of SyGuS problems in the CLIA track of SyGuS-comp. The majority of benchmarks in the CLIA track come from three families of benchmarks: ArraySearch, ArraySum, and ArrayMax. Each of these problems asks for an LIA term that computes a function over a fixed-sized array.

*SyGuS-Imp.* The SyGuS-Imp suite consists of 27 imperative versions of some of the ArraySearch, ArraySum, and ArrayMax family of SyGuS problems. They are a family of SemGuS benchmarks—for an imperative language using while loops, assignment, and conditionals—to perform the same tasks in an imperative setting (similar to the one in Figure 1a). There are 27 benchmarks (for arrays of size two to ten for each family of benchmark).

*FuncImp.* The FuncImp suite consists of 19 simple imperative and functional program-synthesis tasks. This suite of benchmarks consists of 19 SemGuS problems: two using the language from Figure 1; ten using the loop and increment language from Figure 2; and seven using a simple deterministic functional language with match statements (similar to OCaml).

*Boolean.* The Boolean suite consists of 88 SemGuS problems that require synthesizing Boolean formulas in CNF, DNF, and cube format that are logically equivalent to a separate formula that may be in any format.

*Regex.* The Regex suite consists of 51 SemGuS problems that require synthesizing regular expressions for strings with a fixed maximum size (represented using one integer variable per character), 41 of which come from the official SemGuS benchmarks. The remaining 10 are similarly defined, except that the specification requires synthesizing a regular expression for some restricted grammar that is equivalent to a regular expression from a less-restricted grammar.

*ScoreCards.* The ScoreCard suite consists of 30 SemGuS problems that require synthesizing a score-card. Score-cards are used in many areas including sports, games, management, and healthcare to provide an objective overview of performance. A score-card keeps track of tasks performed or objectives achieved, and assigns each a score to create an overall metric [27]. We consider a DSL that represents a score-card as a list of Boolean formulas and arithmetic expressions of the form $\langle b_1, c_1 \rangle, \ldots, \langle b_k, c_k \rangle$ that is effectively interpreted as the imperative program:

$$\text{if } b_1 \text{ then } counter \mathrel{+}= c_1; \ldots; \text{if } b_k \text{ then } counter \mathrel{+}= c_k;$$

Note that while each $c_i$ is additive, $c_i$ may evaluate to a negative value. The goal of each problem is to synthesize a score-card that ensures the counter remains in a bounded region. Specifically, each synthesis problem requires synthesizing an LIA formula for each $b_i$ and an LIA term for each $c_i$ over a set of variables $\bar{v}$, which represent the number of times that a task/objective was achieved. The specification constrains the possible valuations of $\bar{v}$ to be ones for which the overall score is in a bounded region (i.e., $d \leq counter \leq e$ for some LIA terms $d$ and $e$).

*Controllers.* The Controllers suite consists of 40 SemGuS$^\mu$ problems, each of which encodes a reactive-synthesis problem. The first 20 consist of reactive-synthesis problems that require the synthesized (while) program to maintain the controller variables within a safe bound. The last 20 require showing that a reactive program satisfies a linear temporal property (LTL formula): 10 require synthesizing a reactive program for a fixed LTL property; the other 10 require synthesizing an LTL specification for a fixed reactive program. Each of the problems we consider involves proving reactive properties of a while program that never terminates (i.e., of the form `while true do S`). The first suite requires synthesizing a loop body $S$ that ensures that on every iteration (forever) the controller's variables are within a bounded region. For example, we consider problems that model controllers for thermostats, cruise control in cars, and agricultural sprinkler systems. In the thermostat example, the controller must decide when to turn on and off heating and cooling to maintain a specify temperature range. Similarly, the cruise-control example must determine how to appropriately accelerate or brake the car to maintain a given speed range, and the sprinkler system must determine the flow of water to maintain an appropriate range of soil moisture levels.

The final set of problems consider similarly defined programs, except that the specification is now an LTL property. We consider a subset of LTL that includes state predicates (i.e., LIA formulas over the current state of the program); a *next* operator $N$, which requires the formula to hold on the next iteration of the program; an *always* operator $G$, which requires the formula to hold for the current iteration and all future iterations; an *eventually* operator $F$, which requires the formula to hold for some future iteration of the program; and Boolean combinations of these temporal operators. In our experiments, we consider LTL specifications including, "eventually the program reaches a state in which $\varphi$ holds in all future states" ($F(G(\varphi))$) and "it is always the case that the program eventually satisfies $\varphi$" ($G(F(\varphi))$) where $\varphi$ is some LIA predicate over the program's variables.

*PDDL.* The PDDL suite of benchmarks consists of 10 SemGuS$^\mu$ problems encoding problems in the Planning Domain Definition Language (PDDL) [19]. A PDDL problem consists of a set of objects, a set of actions through which the robot can interact with the objects and its environment, a set of predicates the robot can query to observe the state of the world, and a goal (represented as a formula over the predicates) specifying the task the robot should perform. Solving a PDDL problem requires synthesizing a schedule (sequence of actions) that ensures the robot achieves the goal. Similar to the Büchi game shown in Figure 3, each of the problems considers a robot in a grid-like environment to perform a task. In these problems we model a robot that needs to water plants by transporting water from a well to each plant. The set of objects consists of the water well and each plant (each of which can be located by the robot), the robot can move 1 space north, east, south, or west; refill its water (if it's at a well); or water a plant (if its at the plant). Finally, the set of predicates let the robot determine if it has any remaining water and if each plant is sufficiently watered. Each problem is a variant of this problem in which the location of objects, capacity of the robot, and amount of water each plant needs is varied. The goal is for the robot to ensure that each plant is eventually sufficiently watered.

*Games.* The Games suite consists of 20 SemGuS$^\mu$ problems that encode variations of reachability and Büchi games similar to the one shown in Figure 3 using different values for the parameters of the game (e.g., bounded vs. unbounded regions, and whether or not the target is stationary). Büchi games are especially interesting for Muse because the specifications require a semantics that uses both least and greatest fixed-points (in the form of a negated relation).

*Generating Logical Specifications.* For the 141 pre-exisisting SemGuS problems with input-output examples as specifications, we handcrafted a logical specification for each as follows. A majority of the problems included comments that informally defined a logical specification (e.g., "doublex should terminate in a state where y is twice the input value of x"—i.e., "$y' = 2x$"). We used this information to formulate the logical specification that we used in our version of the benchmark (e.g., $\forall x, y, x', y'. Sem_S(doublex, x, y, x', y') \Rightarrow y' = 2x$). When such comments were not available, we produced a logical specification that is satisfied by the expected solution and is consistent with the IO examples. (11 of the 141 SemGuS problems lacked such comments.)

*Benchmark Statistics.* Over all benchmarks, each grammar had at least 1 non-terminal, at most 5 non-terminals, and on average 3 (median 3) non-terminals. Each grammar had at least 1 production, at most 51 productions, and on average 14 (median 12) productions. Each non-terminal could derive at least 1 production, at most 45 productions and on average 5.5 (median 4) productions.

For SemGuS problems, each non-terminal is associated with exactly one semantic relation. For SemGuS, each non-terminal had at least 1 semantic relation, at most 4 semantic relations, and on average 2.1 (median 2) semantic relations. All benchmarks are written in the SemGuS$^\mu$ format, where each production has exactly 1 corresponding semantic rule per semantic relation of the corresponding non-terminal. Each semantic rule had a minimum AST size of 1, a maximum AST size of 524,837, and an average AST size of 1142.7 (median 51.3)

For each SemGuS problem, the semantics is expressed entirely via CHCs. SemGuS$^\mu$ problems typically resulted in a single alternation from a greatest fixed-point to a block of least fixed-points. Some SemGuS$^\mu$ problems (the more complex LTL controller examples) included two alternations: a block of least fixed-points, a greatest fixed-point, and finally a block of least fixed-points.

For all SemGuS and SemGuS$^\mu$ problems, the specification of the desired program had a minimum AST size of 3, a maximum AST size of 7,435, and an average AST size of 197.5 (median 35). Each specification begins with a block of universally quantified variables. The majority of specifications had 0 quantifier alterations, and 20 specifications had a single quantifier alternation. Each specification had a minimum of 0 quantified variables, a maximum of 22 quantified variables, and an average of 7.2 (median 7) quantified variables.

Each program has a minimum AST size of 1, a maximum AST size of 172, and average AST size of 21 (median 8).

## C  PROOFS OF THEOREMS

THEOREM 4.3 (SMT-FORMULA-OF IS SOUND). *For any SemGuS$^\mu$ problem $\mathcal{P} = \langle G = \langle N, \Sigma, S, T, a, \delta \rangle, \langle SEM, [\![\cdot]\!] \rangle, F, \varphi \rangle$ and solution $P$ of $\mathcal{P}$, if $Sem_A$ is non-recursive on $P(f)$ for each occurrence of $Sem_A(f, \overline{x})$ within the specification $\varphi$, then SMT-FORMULA-OF$(\mathcal{P}, P)$ is valid if and only if $P$ is a valid solution of $\mathcal{P}$.*

PROOF. The SMT-FORMULA-OF procedure replaces the definition of each semantic relation appearing in the specification $\varphi$. The process works by performing a fixed-point computation, recursively replacing each occurrence of a semantic relation within a definition with the definition of the semantic relation. That is it replaces each occurrence of $Sem_A(t, \Gamma, \Upsilon)$ with it's definition $\varphi$-of $(Sem_A(t, \Gamma, \Upsilon))$. Clearly, each updates maintains the semantics of the semantic relation of interest. That is, we are simply replacing occurrences of semantic relations with their definitions. Finally, we must show that if $Sem_A$ is non-recursive on $t$, then SMT-FORMULA-OF terminates. Assume not, then there must be some infinite sequences of substitutions that replace a semantic rule with its definition. However, the semantics can only be applied to sub-terms of the program of interest $t$ (or to $t$ itself). If $t$ is strictly decreasing then clearly, SMT-FORMULA-OF must terminate. If instead, there is an infinite sequence of reductions using the same term $t$, then this violates the assumption that $Sem_A$ is non-recursive on $t$. Thus we may conclude SMT-FORMULA-OF terminates.  □

THEOREM 4.5 (CHC-OF IS SOUND.). *For any SemGuS$^\mu$ problem $\mathcal{P} = \langle G = \langle N, \Sigma, T, a, \delta \rangle, \langle SEM, <_{SEM}, [\![\cdot]\!] \rangle, F, \varphi \rangle$ and solution $P$ of $\mathcal{P}$, if for each occurrence of $Sem_A(f, \overline{x})$ within the specification $\varphi$, it appears negatively and the rules defining $Sem_A$ are CHC-like for $P(f)$, then the query returned by CHC-OF$(\mathcal{P}, P)$ is valid if and only if $P$ is a valid solution of $\mathcal{P}$.*

Proof. Let $rules \models \psi$ be the query returned by CHC-OF. We must prove that $rules^{LFP} \models \psi$ if and only if $SEM^{LFP} \models \varphi$. The CHC-OF procedure proceeds by performing a search over the semantics and tree $t$ jointly to compute a set of CHCs that capture the semantics of $t$. On each iteration, CHC-OF picks a semantic relation $Sem_A$ and subtree $t'$, and computes a set of CHCs logically equivalent to the rule defining $Sem_A(t', \Gamma, \Upsilon)$. Then each semantic relation sub-tree pair appearing in the rule defining $Sem_A(t', \Gamma, \Upsilon)$ is added to the queue, ensuring that set of computed CHCs captures the semantics of $t$ (and all it's sub-trees) rather than just the semantics of the root production of $t$. Moreover, you cannot talk about the "root production rule" alone; you have to talk about the root production rule of a specific tree. On each iteration, the CHC rules computed by CHC-OF are computed from the cubes of the disjunctive normal form of the single rule defining the semantic relation $Sem_A$ for the root production rule of the tree $t$. Clearly, these rules are logically equivalent to the single rule they were derived for (i.e. it is easy to show $(a \leftarrow b) \wedge (a \leftarrow c)) \Leftrightarrow (a \leftarrow (b \vee c))$. Thus once the algorithm terminates, we are guaranteed that the resulting rules are logically equivalent to the original semantic relations. The proof of coCHCs proceeds similarly. □

Theorem 4.6 (Verification of Semgus is not reducible to CHC Satisfiability). *There exists a program $t$ and SemGuS problem $Sy$ such that verifying $t$ satisfies $Sy$ cannot be reduced to satisfiability of Constrained Horn Clauses.*

Proof. To prove that some solution $T$ satisfies a SemGuS problem $Sy$, one must prove that the least solution to the semantic rules $SEM_{Sy}$ of $Sy$ satisfies the specification $\varphi_{Sy}$ when the solution $T$ is substituted for each occurence of a synthesis function (i.e. $SEM_{Sy}^{LFP} \models \varphi_{Sy}[f \mapsto T(f) : f \in dom(F_{Sy})]$). By assumption, we know that the semantic rules $SEM_{Sy}$ are represented as a set of CHCs, and the specification $\varphi_{Sy}$ is given as an arbitrary first-order formula with no free variables that may include any of the semantic relations defined by the semantic rules. Specifically, the specification $\varphi_{Sy}$ may contain both negative *and* positive occurrences of semantic relations. Thus in general (and specifically for $\varphi_{comm}$ in eq. (8)), the specification *is not* a valid query formula for checking satisfiability using constrained horn clauses.

Recall that satisfiability problem for CHCs is typically formulated as reachability (i.e., given a set of CHCs and a formula, can the formula be derived from the set of CHC rules?). This formulation is known to be logically equivalent to determining if some interpretation (namely the least interpretation [25]) of the uninterpreted relations satisfies every CHC rule and the given formula when the provided query formula is of the form [9]:

$$\forall \bar{x}. R(x) \Rightarrow \psi$$

where $R$ is an uninterpreted relation defined by the CHC rules, and $\psi$ is a constraint over the variables $\bar{x}$. Note, that there is no way to transform arbitrary first-order formulas (i.e. SemGuS specification) to this form. Thus we may conclude that there are some SemGuS verification problems not reducible to CHC satisfiability. □

Theorem 4.7 (muCLP-OF is sound). *For any SemGuS$^\mu$ problem $\mathcal{P} = \langle G, \langle SEM, <_{SEM}, [\![\cdot]\!]\rangle, F, \varphi\rangle$ and solution $P$ of $\mathcal{P}$, the query returned by muCLP-OF$(\mathcal{P}, P)$ is valid iff $P$ is a valid solution of $\mathcal{P}$.*

Proof. Let $rules \models \psi$ be the query returned by muCLP-OF. We must prove $rules^{FP} \models \psi$ if and only if $SEM^{LFP} \models \varphi[f \mapsto P(f)]$. We proceed to prove by induction on the number of iterations of the main loop of muCLP-OF, that $rules$ are logically equivalent to to rules defining the semantic relations that have been explored previously. Trivially this holds true before entry to the loop. Let $\langle Sem_A, t', fix \rangle$ be the next elements to processed by the main loop of muCLP-OF. The first step is to retrieve the rule defining the $Sem_A$ for the root production rule of $t'$. Next, we optionally dualize the the rule if $fix$ is $\nu$ (i.e. we are interested in computing the dual relation of $Sem_A$). Next, the rule is normalized. This process replaces very negative occurrence of a semantic relation with it's dual relation (i.e. $Sem_{A'}^{\neg}$). Each of these transformations preserves the semantic relation's interpretation. Then finally, every positive occurring semantic relation $Sem_{A_j}$ is added to the queue to compute its semantics, and every semantic relation appearing negatively is added to the queue to compute it's dual semantics. Thus, in some future iteration, each semantic relation of interest will be processed and its fixed-point equation added to the set of rules. Once, MUCLP-OF, terminates it does so with a set of rules *rules* that captures the semantic relation appearing in $\varphi$ (and the semantic relations needed to define the semantics for each sub-tree of $t$). The final step of the process, orders the set of fixed-point equations based on the ordering $<_{SEM}$ (i.e., if $Sem_A < Sem_B$, then $Sem_A$ is evaluated before $Sem_B$ within the fixed-point computation $rules^{FP}$). Since $\psi = Norm(\varphi)$, it's clear that $rules^{FP} \models \psi$ if and only if $SEM^{LFP} \models \varphi$. □

THEOREM 4.8 (SEMGUS$^\mu$ SEMANTICS AND $\mu$CLP ARE EQUALLY EXPRESSIVE). *For every $\mu$CLP query $\langle \varphi, preds \rangle$, there is some SemGuS problem $\mathcal{P}$ and solution $P \in L(G_\mathcal{P})$ such that $\langle \varphi, preds \rangle$ is valid if and only if $P$ is a valid solution to $\mathcal{P}$.*

PROOF. Let $X_0(\bar{x}_0) =_{fix_0} = \varphi_0; \ldots; X_n(\bar{x}_n) =_{fix_n} \varphi_n$ be the sequence of predicates making up *pred*.

Consider the following grammar $A := \top$, that consists of a single nonterminal $A$ with a single production rule $\top$. That is the the language of $A$ consists of a single word $L(A) = \{\top\}$. First, define $Y_i$ to be $X_i$ if $fix_i$ is $\mu$ and $X_i^{\neg}$ otherwise, and similarly, $\psi_i$ to be $\varphi_i$ if $fix_i$ is $\mu$ and $\neg\varphi_i$ otherwise.

Let $SEM_A = \{Y_0, \ldots, Y_n\}$ be the set of newly introduced predicate relations. For each $Y_i$ define $[\![\top]\!]_{Y_i}$ to be the rule $Y_i(\top, \bar{x}_i) \leftarrow \psi_i[m]$ where $m$ maps every occurrence of $X_j(\bar{x}_j)$ in $\psi_i$ to $Y_j(t_A, \bar{x}_j)$ if $fix_j$ is $\mu$ and $\neg Y_j(t_A, \bar{x}_j)$ otherwise ($t_A$ is a variable representing elements of $L(A)$). Similarly, define $\varphi_{spec} = \varphi[m]$ to be the constraint of the muclp query using the same substitution $m$. Finally, define $\mathcal{P}$ to be the SemGuS$^\mu$ problem defined by the grammar for $A$, semantics $\langle SEM_A, \{Y_i < Y_j : i < j\}, [\![]\!] \rangle$, specification, and set of synthesis functions $F = \{\langle t_A, A \rangle\}$.

In order to prove that $pred^{FP} \models \varphi$ if and only if $SEM_A^{LFP} \models \varphi_{spec}$, we prove that for each predicate $X_i$ that if $fix_i$ is $\mu$ then the fixpoint of $X_i$ is the least fixpoint of $Y_i$ ($X_i^{FP} = Y_i^{LFP}$); otherwise, that the fixpoint of $X_i$ is the dual of the least fixpoint of $Y_i$ ($X_i^{FP} = \neg Y_i^{LFP}$). We proceed by induction on $n$ the number of rules defining the $\mu$CLP query. First, consider the case when there are no rules. Trivially, the claim holds. Now, assume that for each $i > 0$ that the claim holds for $X_i$ and $Y_i$. We now proceed to prove the case for $X_0$ and $Y_0$. Consider the case when $fix_0$ is $\mu$. By definition $X_i^{FP} = X_i^{LFP}$, $X_i = Y_i$ and $[\![\top]\!]_{y_i} = Y_i(\top, \bar{x}_i) \leftarrow \varphi_i[m]$. Both $X_i$ and $Y_i$ are defined as least fixed-points. Additionally $\varphi_i[m]$ is identical to $\varphi_i$ except every occurrence of $X_j$ has been replace with either $Y_j$ if $fix_j$ is $\mu$ or by $\neg Y_j$ if $fix_j$ is $\nu$. In either case, it is clear from the inductive hypothesis that the substitution preserves fixed-points. Since, both rules are defined as least fixed-points, with logically equivalent definitions, we can conclude that $X_i^{FP}$ and $Y_i^{LFP}$ are equivalent. Next, we consider the case when $fix_0$ is $\nu$. Then by definition, $Y_0 = X_0^{\neg}$ and $[\![\top]\!]_{Y_0}$ is $Y_0(\top, \bar{x}_0) \leftarrow \neg\varphi_0[m]$. Similar to the previous case, we may use the IH to assume that for each $X_j \neq X_0$ the substitution substitutes equivalent terms. We note, that the rule defining $Y_0$ is dual to the rule defining $X_0$. Thus, we can may conclude this case and have proved that if $fix_i$ is $\mu$ then the fixpoint of $X_i$ is the least fixpoint of $Y_i$ and otherwise $X_i$ is the dual fixpoint of $Y_i$. □

THEOREM 4.9 (SEMGUS AND $\mu$CLP ARE NOT EQUALLY EXPRESSIVE). *Verifying solutions to SemGuS problems can be encoded within a fragment of $\mu$CLP that uses at most one alternation between greatest and least fixed-points.*

PROOF. As stated in Theorem 4.8, every SemGuS$^\mu$ verification problem is expressible as a $\mu$CLP query and vice versa. Thus, verification of solutions to SemGuS$^\mu$ problems requires the full generality of the $\mu$CLP calculus. Since every SemGuS problem is equivalently representable as a SemGuS$^\mu$ problem, clearly verification of SemGuS problems can be reduced to validity of a $\mu$CLP query. However, verification of solutions to SemGuS problems do not require the full generality of the $\mu$CLP calculus. Specifically, the encoding described in MUCLP-OF will result in a $\mu$CLP formula that (at most) contains equations describing the semantic and dual semantic relations with no interaction between the two (i.e. does not require any interaction between greatest and least fixed-points). Thus we may conclude the fact that verification of solutions to SemGuS$^\mu$ problems in general require a more expressive logical encoding than the encoding of verification of solutions to SemGuS problems. □

THEOREM 5.3 (REIFICATION IS SOUND). *Given a formula $\varphi$ and a set, rules, of semantic rules, let $\psi$ be the formula in which every occurrence of $Sem_A(t, \overline{x})$ is replaced by the reified semantic relation $Sem_A^t(\overline{x})$, and rules$^{reify}$ is the conjunction of the reified semantic rules produced by REIFY(rules, $Sem_A$, $t$) for each $Sem_A(t, \overline{x})$ appearing in $\varphi$. The constraint $\varphi$ is valid under the original semantic rules rules if and only if $\psi$ is valid under the reified semantic rules: rules $\models \varphi \Leftrightarrow$ rules$^{reify} \models \psi$.*

PROOF. We begin by induction on the height of the tree $t$. The case when $t$ has height 0 is trivial. Next, assume for any subtree $t'$ of $t$ that $Sem_A(t', \Gamma', \Upsilon')$ is logically equivalent to $Sem_A^{t'}(\Gamma', \Upsilon')$. REIFY computes the rule $Sem_A^t(\Gamma, \Upsilon) \leftarrow \varphi[m]$ where $m$ is a map substituting every occurrence of $Sem_{A_j}(t_j, \Gamma_j, \Upsilon_j)$ with $Sem_{A_j}^{t_j}(\Gamma_j, \Upsilon_j)$. Either $t_j$ is a subtree of $t$ or $t_j = t$. In the first case, the inductive hypothesis may be used to show that the substitution preserve logical equivalence of the two rules. Otherwise, if $t_j = t$, we use coinduction to show that the possibly infinite cycle of semantic rules with program terms are logically equivalent to their reified version. The argument holds that if the property holds, then it will continue to hold regardless of how the semantic rules are defined. We may then conclude that the reified rules and original rules are logically equivalent. □